\documentclass[fleqn,usenatbib]{mnras}

\usepackage{newtxtext,newtxmath,subcaption}

\usepackage[T1]{fontenc}

\DeclareRobustCommand{\VAN}[3]{#2}
\let\VANthebibliography\thebibliography
\def\thebibliography{\DeclareRobustCommand{\VAN}[3]{##3}\VANthebibliography}

\usepackage{graphicx}	
\usepackage{amsmath}	

\title[Blueshifted absorbers]{Blueshifted lines from the inner accretion disc's rotation can explain quasar absorption ``forests''}

\author[A. M. Hankla et al.]{
Amelia M. Hankla,$^{1,2}$\thanks{E-mail: \href{mailto:lia.hankla.astro@gmail.com}{lia.hankla.astro@gmail.com}, \href{mailto:ahankla@princeton.edu}{ahankla@princeton.edu}}%
\thanks{NASA Hubble Einstein Fellow}
Fergus J. E. Baker,$^{3}$
Daniel R. Wilkins$^{4}$
and Andrew C. Fabian$^{5}$\thanks{E-mail: \href{mailto:acf@ast.cam.ac.uk}{acf@ast.cam.ac.uk}}
\\
$^{1}$ Department of Astronomy, University of Maryland, 4296 Stadium Dr Ste 1113, College Park, MD 20742, USA\\
$^{2}$ Department of Astrophysical Sciences, Princeton University, 4 Ivy Lane, Princeton, NJ 08544, USA\\
$^{3}$School of Mathematics, Statistics and Physics, Newcastle University, Herschel Building, Newcastle upon Tyne, NE1 7RU, UK\\
$^{4}$Department of Astronomy, The Ohio State University, 4055 McPherson Laboratory, 140 W 18th Ave, Columbus, OH 43210, USA\\
$^{5}$Institute of Astronomy, University of Cambridge, Madingley Road, Cambridge CB3 0HA, UK
}

\date{Accepted 2026 November 17. Received 2026 November 16; in original form 2026 April 10}

\pubyear{\the\year{}}

\begin{document}
\label{firstpage}
\pagerange{\pageref{firstpage}--\pageref{lastpage}}
\maketitle

\begin{abstract}
Recent XRISM observations of active galactic nuclei (AGN) such as PDS 456 have revealed ``forests'' of absorption lines best modeled by five distinct absorption zones with varying large blueshifts. 
We propose a model in which these relativistic blueshifts originate from the motion of the accretion disc itself, rather than from a clumpy super-Eddington outflow at hundreds of gravitational radii $r_g\equiv GM/c^2$. 
We demonstrate that thin fragments of absorbing material lying just above the accretion disc at varying radii can produce the observed energy shifts and separations of the absorption zones.
In this model, the PDS 456 X-ray transmission spectrum is well reproduced by fragments of azimuthal rings with radial widths $\Delta r\lesssim1r_g$ at locations between the black hole's innermost stable circular orbit (ISCO) and $\approx15r_g$.
This model suggests that the absorption forests seen in XRISM observations can probe the surface structure of the innermost ($\lesssim15r_g$) regions of quasar accretion discs and predicts the presence of redshifted absorbers.
\end{abstract}




\section{Introduction}
Since its launch in 2023, the X-Ray Imaging and Spectroscopy Mission (XRISM;~\citealt{xrism}) has used its high spectral resolution in the X-rays to reveal ``forests'' of absorption features in at least two quasars: PDS 456~\citep{xu2025, pds456} and PG 1211+143~\citep{mizumoto2026, reeves2026}.
These absorption forests comprise numerous narrow absorption lines with widths $\sim0.2$ keV.
Photoionization modelling suggests the several absorbing components -- five, in the case of PDS 456~\citep{pds456} -- each with distinct blueshifts.
The dominant hydrogen- and helium-like iron (Fe XXV and Fe XXVI) lines shift from their rest-frame energies of 6.67 and 6.97~keV, respectively, to between 8 and 9~keV. 

The prevailing model for these absorption forests proposes that clumps of absorbing gas located approximately 200 to 500 $r_g$ from the black hole absorb the X-ray emission from the central corona while moving away from the black hole in an ultrafast outflow (UFO) with speeds of 0.2 to $0.3c$~\citep{pds456, mizumoto2026}.
In PDS 456, the outflow is stratified in velocity with five non-overlapping clumps along the line of sight, implying an overall volume filling fraction of 0.1 to 0.3.
The resulting wind kinetic power exceeds the estimated Eddington luminosity of the black hole by about a factor of 2 and requires a covering fraction close to or above 100\%~\citep{pds456}, suggesting that such a wind would strongly couple to the
surrounding interstellar medium and affect the subsequent evolution of the galaxy (see~\citealt{fabian2012} for a review).
Launching radii of $\sim10r_g$ have been inferred for the fastest components in PDS 456 by setting the velocity equal to the local escape speed~\citep{reeves2018, boissay2019}, consistent with clumps travelling outward by the time they are observed.
Although these estimates imply gas located in the inner disc radii, they do not establish that the gas is unbound and as such do not explicitly rule out an alternative origin for the relativistic velocity of the gas.

To robustly test the prevailing UFO model and its implications for AGN feedback, this work explores whether these absorption features could instead originate within the accretion disc itself, negating the requirement for an outflow at sub-pc scales. 
In our model, dense, absorbing gas magnetically connected to the disc surface (analogous to solar prominences) moves with the relativistic Keplerian velocity of the inner disc. 
Velocity stratification arises naturally from the $r^{-1/2}$ dependence of the disc azimuthal velocity, rather than changes in UFO velocity.
Similar inner-disc structures such as density spiral waves~\citep{papaloizou1984} or cold patches in the accretion disc~\citep{armitage2003} have been invoked to explain varying or transient iron K$\alpha$ lines in other AGN~\citep{iwasawa2004, marinucci2020}.
A disc origin for this absorption forest provides a powerful tool for probing inner disc properties such as density and ionization and does not preclude the existence of larger-scale winds, such as the kiloparsec-scale molecular and ionized outflows seen in PDS 456~\citep{Bischetti2019, Travascio2024, Seebeck2024}.
Constraining these properties will better inform subsequent models for magnetic flux advection and jet launching under the assumption of no significant outflow from the inner disc. 

Building upon previous works exploring disc-tied absorption~\citep{gallo2011, gallo2013, fabian2020}, we demonstrate how this framework can produce multiple absorption zones and constrain the required physical conditions.
After detailing our model (Sec.~\ref{sec:model}) and ray-tracing methods (Sec.~\ref{sec:methods}), we explore the line profiles of a single absorbing ring, characterizing the dependence on ring radial width (Sec.~\ref{sec:lineprofile}), maximum energy shift, and full width at half-maximum (Sec.~\ref{sec:maxg}). 
We then apply our model to PDS 456 and constrain the absorber and system parameters such as spin and inclination angle necessary to reproduce its absorption forest (Sec.~\ref{sec:pds456}). 
We conclude by comparing to other methods that provide inclination angles and discuss implications for variability.

\begin{figure*}
    \centering
    \begin{subfigure}[t]{0.4\textwidth}
        \centering
        \includegraphics[width=0.95\textwidth]{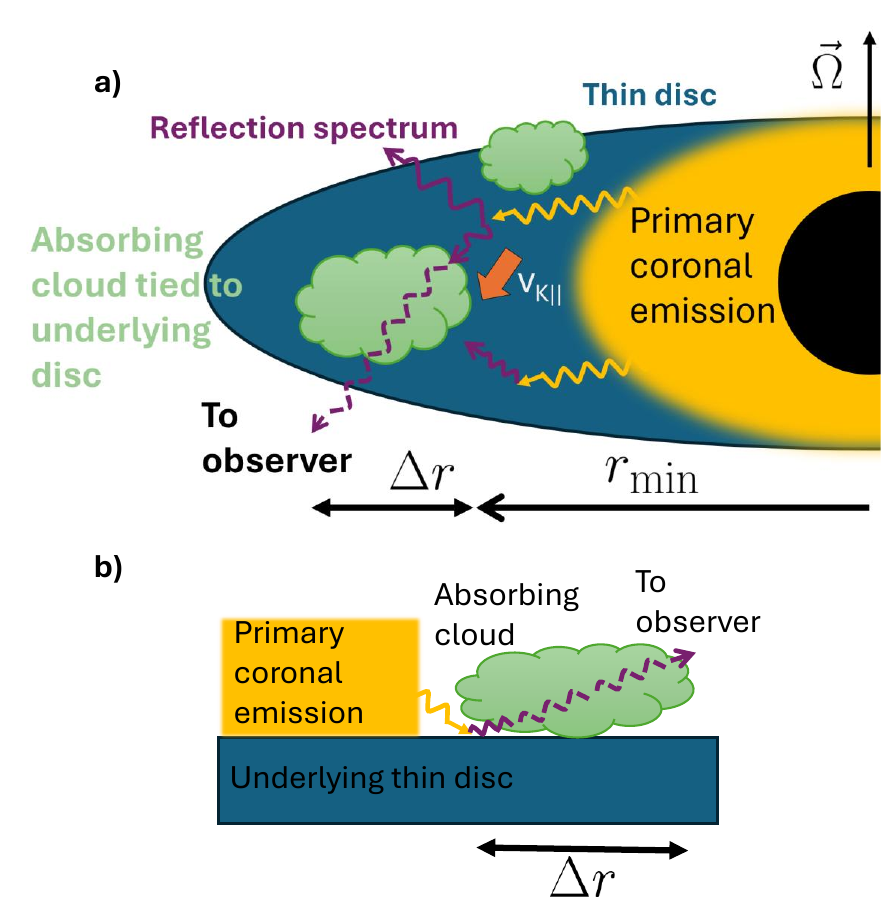}
    \end{subfigure}%
    ~ 
    \begin{subfigure}[t]{0.6\textwidth}
        \centering
        \includegraphics[width=0.95\textwidth]{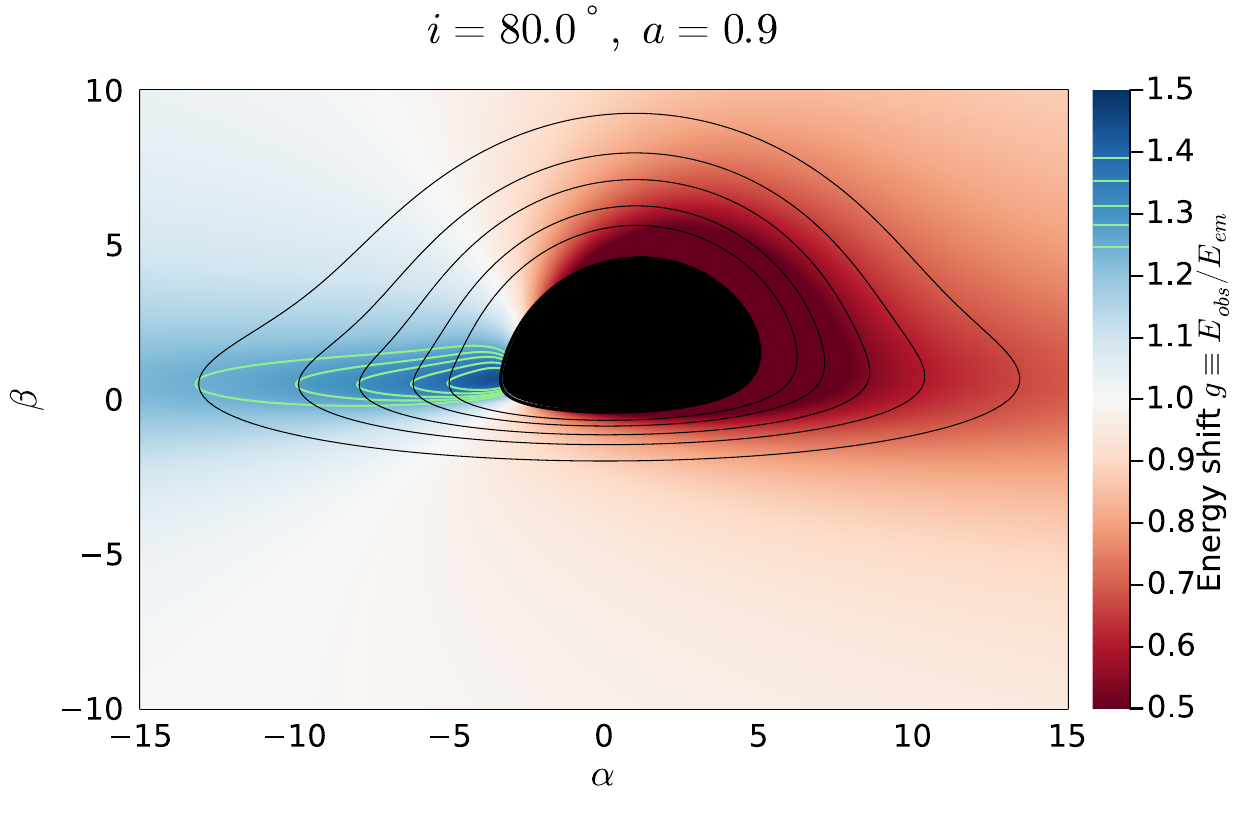}
    \end{subfigure}
    \caption{Left a): Schematic of the geometry proposed to produce absorption lines. A cloud of dense gas (green) sits above a thin accretion disc (blue) with a Keplerian rotation velocity profile (orange arrow). The X-ray corona illuminates the disc with a given emissivity profile, reflects off the disc to produce a reflection spectrum (purple) that passes through the dense gas and to the observer. The cloud starts at radius $r_{\rm min}$ and has a width $\Delta r$, both measured in $r_g$. The disc and black hole angular momentum vector $\vec \Omega$ are aligned, and the observer is inclined with angle $i$ relative to $\vec\Omega$ such that the Keplerian velocity projects along the line-of-sight ($v_{K\parallel})$.
    Left b): A cross-section through the side of the disc, with Keplerian velocity out of the page. The absorbing cloud could be magnetically tied to the disc through a coronal prominence.
    Right: Ray-traced example of the left schematic with five absorbing rings whose locations produce energy shifts corresponding to PDS 456 values (see Table~\ref{tab:absorbers}) in green contours. Dark lines indicate where rings at the radii given in Table~\ref{tab:absorbers} would appear, with the inner black part showing the projection of the innermost stable orbit. } \label{fig:schematic}
\end{figure*}
\section{Model: absorption from gas tied to disc orbital motion} \label{sec:model}
Following standard reflection models, we assume that a compact X-ray source (the ``corona''; yellow in Fig.~\ref{fig:schematic} left panel) sits near the black hole and illuminates a thin accretion disc described by~\citet{ss73} and illustrated in blue in Fig.~\ref{fig:schematic}.
For a black hole mass of $\sim10^9M_\odot$, the disc temperature at $10r_g$ is $\approx5\times10^4$K~\citep{ss73}, such that photoionization by the X-ray source determines the ionization state of iron atoms above the disc.

In our framework, the observer views the X-ray spectrum reflected off the disc (purple in Fig.~\ref{fig:schematic} left panel) through a cloud of absorbing dense gas orbiting with Keplerian velocity $v_K(r_{\rm min})$ just above the disc (green in Fig.~\ref{fig:schematic} left panel).
By assuming no significant outflow, this model tests whether observational signatures often attributed to a relativistic wind could instead arise from the inner accretion disc's Keplerian velocity, which reaches speeds of $0.5c$ relative to the local zero angular momentum observer.
As illustrated in Fig.~\ref{fig:schematic}, the brightest part of the illuminated disc corresponds to its approaching side due to relativistic beaming of the emission in the material's direction of motion.
Therefore, the strongest absorption will be from approaching material and will appear blueshifted.
Although we consider axisymmetric absorbers in an infinitely vertically thin annulus located in the midplane around the spinning black hole for simplicity, the azimuthal geometry of the absorbers does not affect the line shape significantly, as discussed in Sec.~\ref{ssec:shape}.
We define $r_{\rm min}$ as the distance from the black hole to where the absorbing annulus begins, and $\Delta r$ as the radial width of the absorbing annulus.

The black hole's dimensionless spin $a$ has an angular momentum vector $\vec \Omega$ that is inclined with an angle $i$ to the observer, and we assume that the disc motion is in the plane perpendicular to $\vec\Omega$. 
To illustrate the distribution of absorbers, Fig.~\ref{fig:schematic}'s right panel shows a ray-traced image of an example configuration that could produce the absorption spectrum in PDS 456 (see Sec.~\ref{sec:pds456}). 

\section{Methods} \label{sec:methods}
To calculate the absorption features from dense gas, we ray trace photon trajectories in Kerr spacetime using \texttt{GRADUS.JL}~\citep{gradus}\footnote{\texttt{GRADUS.JL} is publicly available at \url{https://codeberg.org/astro-group/Gradus.jl}.}.
We use the software to calculate Cunningham's transfer functions to integrate line profiles~\citep{cunningham1975,dauser2010}, and to determine the energy shift $g$ along a photon geodesic, defined as
\begin{equation}
    g=\frac{E_{\rm obs}}{E_{\rm em}},
\end{equation}
where $E_{\rm obs}$ is the photon energy as measured by a distant observer, and $E_{\rm em}$ the photon energy as measured locally by the emitter. 
This $E_{\rm em}$ corresponds to the rest-frame line energy.
The energy shift includes both contributions from the gravitational redshift arising from the spacetime curvature close to the black hole, as well as the (special relativistic) blue- and redshift of the Doppler effect due to the Keplerian motion of the thin accretion disc, and thus also of the absorbing gas.
We self-consistently calculate axis-symmetric emissivity profiles for the reflected emission of the disc due to illumination by the corona~\citep[e.g.][]{wilkins2012}, and specify the assumed illuminating geometry as needed later in the text.

We treat the absorption as an effectively inverted disc emission line.
Assuming a uniform intrinsic opacity across the absorbing cloud, the contribution of any local disc patch to the shifted absorption profile is proportional to the intensity of its backlighting continuum.
We therefore compute the relativistically blurred emission profile and subtract it from the (assumed uniform) continuum to generate the transmission spectrum.
Consequently, when isolating these kinematic effects, several figures in this work display the profiles as positive features rather than absorption troughs.

\begin{figure}
    \centering
    \includegraphics[width=0.95\linewidth]{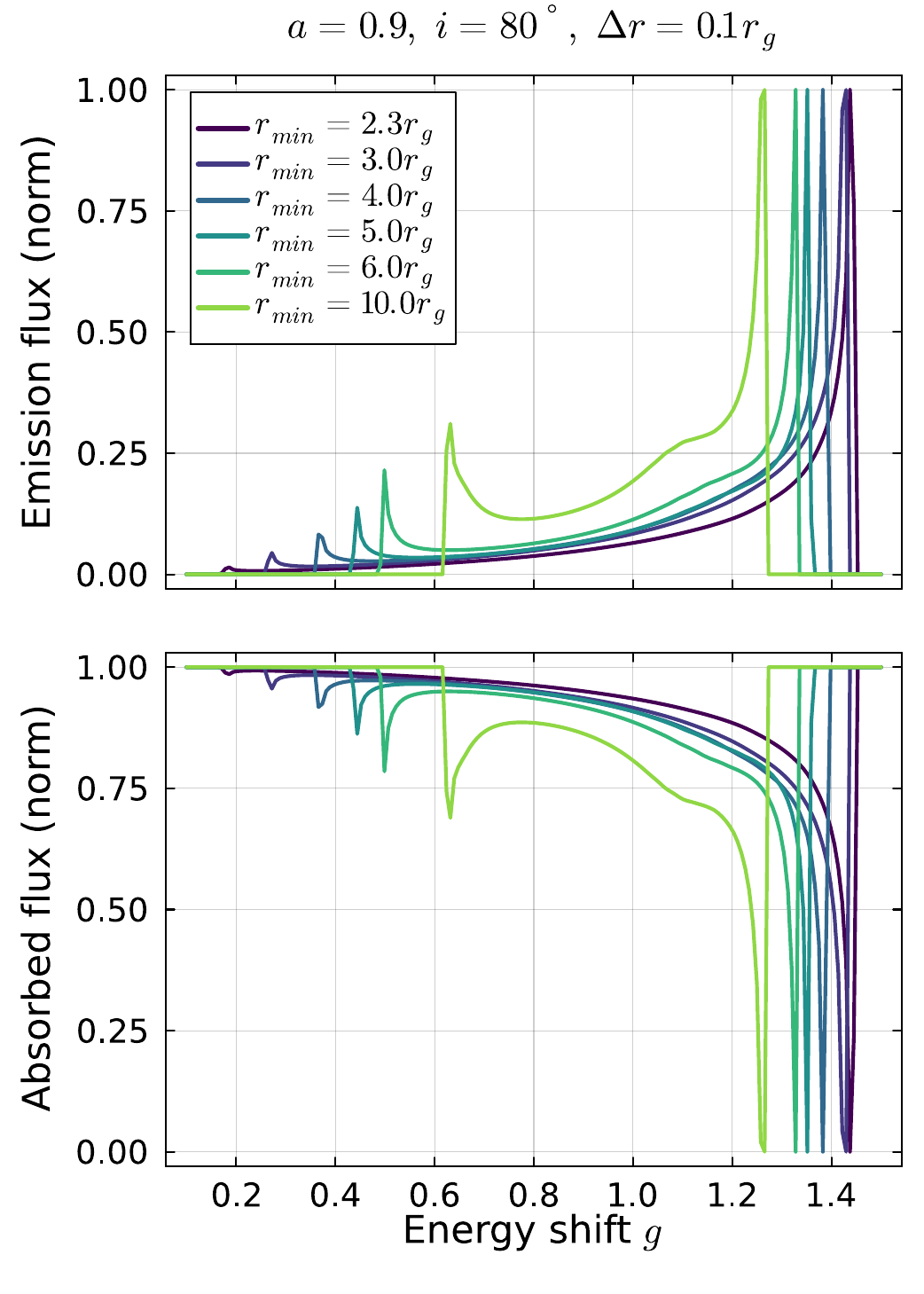}
    \caption{Emitting (top) and absorbing (bottom) rings with velocities tied to the Keplerian disc can produce narrow spectral peaks for a variety of locations $r_{\rm min}$. Fluxes are normalized to their maximum (minimum) value for ease of comparison. The bottom plot shows the lines in the top plot subtracted from 1.}
    \label{fig:varyrmin}
\end{figure}

\begin{figure*}
    \centering
    \includegraphics[width=0.95\linewidth]{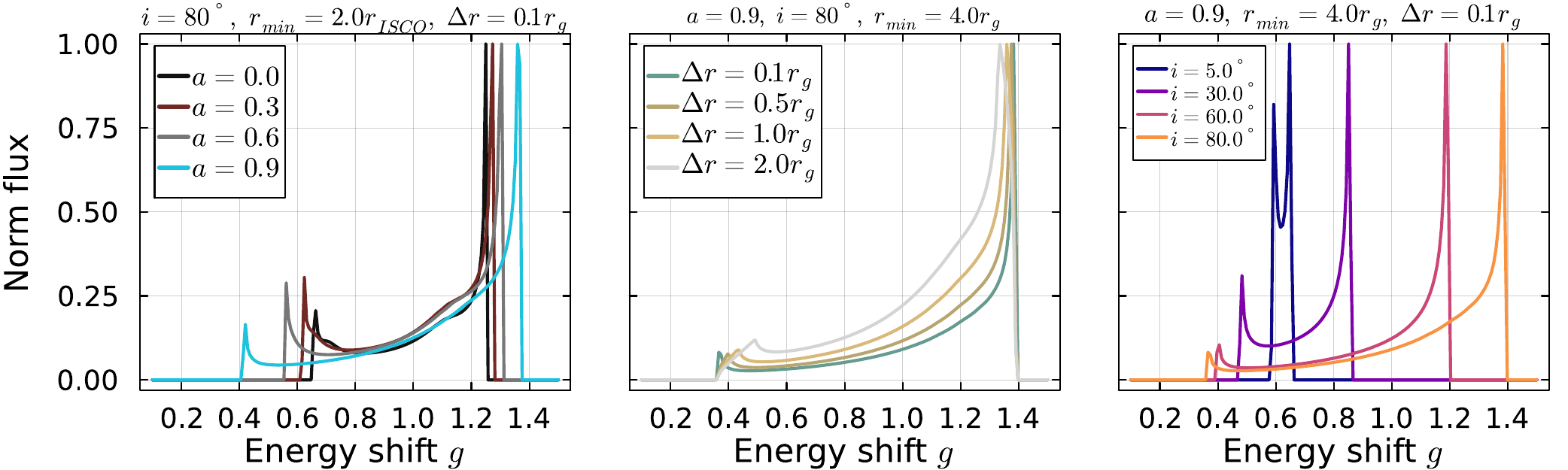}
    \caption{Line emission dependence on black hole spin (first panel), emitting ring radial extent $\Delta r$ (second panel), and observer inclination angle $i$ with respect to the black hole spin axis (third panel). Fluxes are normalized to their maximum value for ease of comparison. }
    \label{fig:vars}
\end{figure*}
\section{Line Profile Dependence on Parameters} \label{sec:lineprofile}
In this section, we consider a single emission line and explore the dependence on the model parameters.
Fig.~\ref{fig:varyrmin}'s top panel shows how the line emission (and thus absorption; bottom panel) varies depending on the radial location of the emitting ring for fixed emitter width. 
As the ring moves closer to the black hole, the peak energy shift increases as well, because the Keplerian velocity of the disc projected along the line of sight changes as $r^{-1/2}$.
The red peak of the line profile shifts to lower energies as the ring moves closer to the black hole, such that emerging photons must climb deeper out of the gravitational well.
At large radii, additional bumps are visible due to light-bending effects, as discussed in~\citet{hameury1994}.

Other parameters include black hole spin, emitter width, and observer inclination angle.
We show the dependence of the line profile on each of these variables in Fig.~\ref{fig:vars}.
This figure normalizes each individual line profile to its maximum flux for better comparison across different parameters.
The unnormalized profile will have different fluxes at the blue peak for different energy shifts since photon conservation changes the emitted-to-observed intensity ratio by a factor of $g^3$.

The first panel in Fig.~\ref{fig:vars} shows how the line profile varies as a function of spin for a fixed ring width, while fixing the ring's location at $2r_{\rm ISCO}$, where $r_{\rm ISCO}$ denotes the radial location of the ISCO. 
As the spin increases, the ISCO moves to smaller radii, resulting in escaping photons sampling higher redshifts (lower $g$).
The red wing of the line therefore shifts to lower values of energy shift $g$ with higher spin.
The dependence of this red wing on the ISCO and hence spin is the cornerstone of using iron line profiles to measure black hole spin~\citep[e.g.][]{reynolds2021}.

The second panel in Fig.~\ref{fig:vars} shows the line profile variation with its radial width $\Delta r$, for fixed black hole spin, inclination angle, and initial position.
As the ring size increases, a wider range of energy shifts is sampled, broadening both the blue and red peaks. 
Because the parts of the ring at $r_{\rm min}+\Delta r$ experience less redshift than those at $r_{\rm min}$, they are also brighter, leading to a red peak that tilts inward (also seen in~\citealt{hameury1994, gates2025}).

The third panel in Fig.~\ref{fig:vars} shows how the two peaks separate based on the observer inclination angle.
For a ring viewed along the black hole spin axis ($i\approx5^\circ$), the emission is nearly a delta function redshifted by the gravitational potential of the black hole. 
As the observer inclination angle increases, the Doppler beaming from the ring edge travelling towards/away from the observer leads to shifting the peak to higher/lower energies.
Due to the conservation of photon number, the blueshifted peak also increases in intensity as $g^3$, while the redshifted peak decreases in intensity. 

Far from the black hole, the dependence on black hole spin and inclination angle is well-described by the formula given in~\citet{hameury1994} Eq. 16.

\section{Constraints from Line Absorption Properties} \label{sec:maxg}
\subsection{Inclination Angle and Spin Constraints From Maximum Energy Shift}
The maximum observed energy shift provides a lower limit on the inclination angle of the observer relative to the black hole spin axis. 
Higher energy shifts typically require larger inclination angles to provide the Doppler boosting necessary to push the line emission to larger energies (seen in Fig.~\ref{fig:vars} right panel). 
To determine the radius at which absorption at a given energy shift $g$ can occur, we now explore the maximum energy shift from a line.
Because a line's maximum flux occurs close to its largest blueshift (Fig.~\ref{fig:vars}), the `maximum energy shift' discussed in this section refers to the energy shift at the line's peak emission flux.

\begin{figure}
    \centering
    \includegraphics[width=0.95\linewidth]{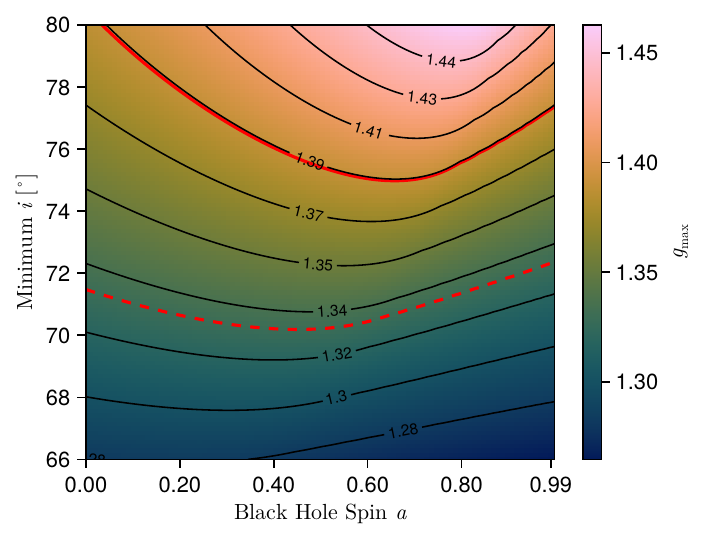}
    \caption{The largest observed energy shift constrains the inclination angle of the accretion disc. For $g_{\max}=1.39$ as seen in PDS 456 assuming an Fe XXV line (Table~\ref{tab:absorbers}), the inner accretion disc around black holes of any spin can produce sufficiently high energy shifts as long as the inclination angle is $\gtrsim75^\circ$ (red solid contour). Lower maximum energy shifts provide weaker constraints on the system inclination angle (e.g. $g_{\rm max}=1.33$, red dashed contour). }
    \label{fig:incconstraint}
\end{figure}

Fig.~\ref{fig:incconstraint} shows the inclination angle constraints provided by a sample of maximum energy shifts, including the maximum observed from PDS 456 assuming an Fe XXV line (solid contour) or Fe XXVI line (dashed contour); see Table~\ref{tab:absorbers} and Sec.~\ref{sec:pds456}. 
For a given maximum energy shift, Doppler boosting by the circular orbit's velocity dominates the inclination constraints at low spins $a\lesssim0.6$.
As the spin increases from $a=0$, the ISCO moves inward while remaining sufficiently far from the event horizon that the impact of gravitational redshift stays relatively constant. 
Gas in circular orbits at the ISCO therefore experiences a higher Lorentz factor, lowering the inclination angle needed to produce a given energy shift.
At higher spins $a\gtrsim0.6$, gas at the ISCO experiences a deep gravitational well, depressing the achievable energy shifts and necessitating higher inclination angles and thus Doppler boosts to produce the observed maximum. 

For an energy shift of 1.39, the balance between Doppler beaming and gravitational redshift occurs around $a\approx0.6$, setting the minimum inclination angle over all spins at $75^\circ$.
For lower observed energy shifts, Doppler beaming can more easily produce the energy shifts regardless of inclination angle, thereby lowering the constraints overall.
The minimum inclination moves to smaller spins for small energy shifts since larger radii can produce the same energy shift and are less affected by movement of the ISCO with spin, so the effects of gravitational redshift become important at lower spins.
Larger energy shifts start to rule out smaller spins, since the maximum energy shift produced by gas at the ISCO viewed edge-on cannot reach energy shifts $\gtrsim1.4$.
These values are consistent with~\citet{gates2020, gates2025}.
Note that if the PDS 456 absorber with the highest energy (9.27 keV;~\citealt{pds456}) comes from Fe XXVI instead of Fe XXV, the largest required energy shift drops to 1.33 instead of 1.39, which softens the constraints on inclination angle (red dashed line; Fig.~\ref{fig:incconstraint}).

\begin{figure*}
    \centering
    \includegraphics[width=0.9\linewidth]{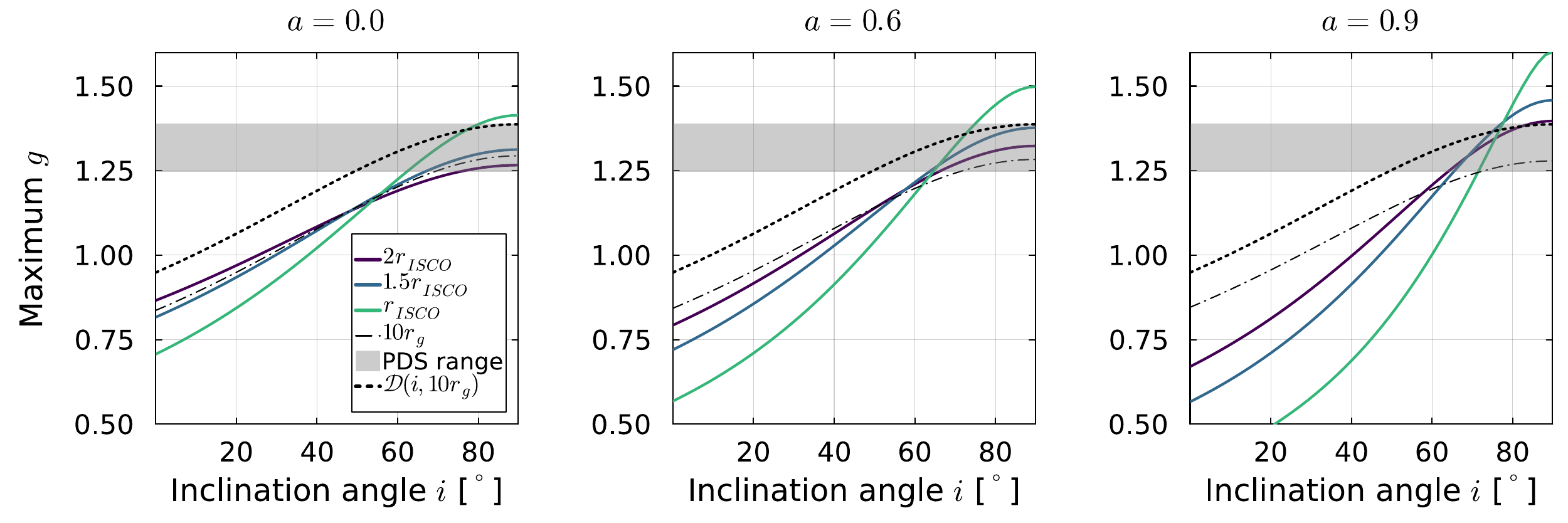}
    \caption{Orbiting absorbers with high observed energy shifts probe the inner $10r_g$ of the accretion disc. For three different spins, solid lines show the maximum energy shift originating from a ring at radius $r_{\rm ISCO}$, $1.5r_{\rm ISCO}$, and $2r_{\rm ISCO}$. The black dash-dot line shows the maximum energy shift originating from $10r_g$, while the black dotted line shows the Doppler factor pure inclination angle dependence for fixed velocity. The gray shaded horizontal region shows the energy shifts observed in PDS 456. }
    \label{fig:maxg}
\end{figure*}

To explore how far from the black hole an absorber must be to produce the energy shifts highlighted above, Fig.~\ref{fig:maxg} plots the maximum energy shift produced by a ring at a given location.
Because the maximum energy shift comes from the smallest radius present in the ring, Fig.~\ref{fig:maxg} plots the maximum energy shift for a given infinitely thin ring location, setting $\Delta r=0$.
Fig.~\ref{fig:maxg} shows that energy shifts between 1.2 and 1.5 easily originate from absorbers located between the ISCO and about $10r_g$. 
Therefore, these energy shifts probe the innermost region of the accretion disc. 

This figure also demonstrates the mechanisms discussed above that constrain inclination angle with spin.
The maximum $g$ produced by a ring at the ISCO (green line) at the highest inclination angles decreases with increasing spin due to the shrinking ISCO. 
For $a=0$, the highest producible energy shift by gas in the disc is right around 1.4, whereas it moves to 1.6 for $a=0.9$. 
Far from the black hole, the dependence on inclination angle is due almost exclusively to Doppler beaming, whose energy shift goes as the Doppler factor $\mathcal{D}=\sqrt{1-\beta^2}/(1-\beta\sin i)$, where $\beta=v/c$. Note that $i$ is the angle between spin and observer, which is $90^\circ$ offset from the angle between gas velocity and observer. 
Fixing $\beta=v_K(10r_g)/c=1/\sqrt{10}$ as the Keplerian orbital velocity at $10r_g$, all the energy shift's dependence is in the inclination angle.
The black dotted line shows the resulting Doppler factor's dependence on inclination angle, which remains constant across spins. 
At lower inclination angles, gravitational redshift dominates the energy shift, causing all emission from within $\approx10r_g$ to have $g<1$.
These results are in accordance with previous findings~\citep{hameury1994, gates2020}.

For non-spinning black holes (Fig.~\ref{fig:maxg} left) at high inclination angle, absorbers within $2r_{\rm ISCO}$ can produce the energy shifts seen in PDS 456.
Spinning black holes (center and right panel) can produce energy shifts larger than the PDS 456 range for radii within $\approx2r_{\rm ISCO}$, thus limiting the absorber location to radii outside the ISCO or suggesting lower inclination angles.

\subsection{Constraints on Number of Absorbers from Line Absorption Red/Blue Peak Separation}
\begin{figure}
    \centering
    \includegraphics[width=0.9\linewidth]{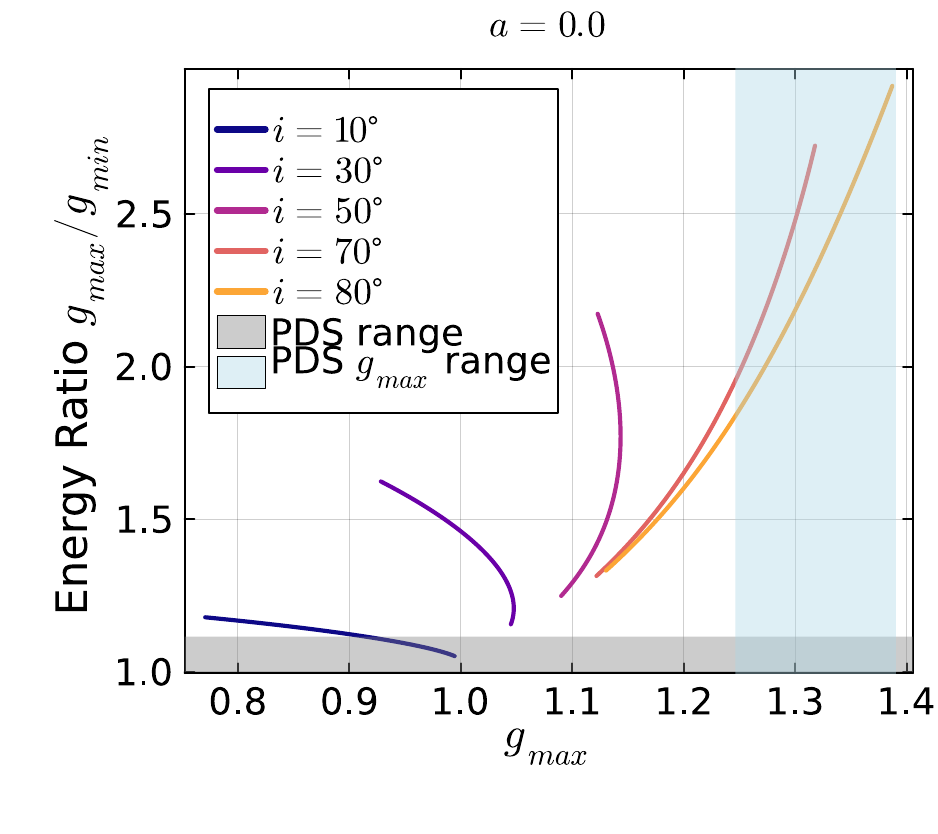}
    \caption{The energy ratio between the blue peak with energy shift $g_{\rm max}$ and the red peak with energy shift  $g_{\rm min}$ as a function of $g_{\rm max}$ demonstrates that PDS 456 must have five separate absorbing zones. Constraints from PDS 456 are given by colored vertical and horizontal spans.}
    \label{fig:gratio}
\end{figure}

The ``forest'' of absorption lines displayed by PDS 456 suggests the presence of multiple absorbers.
Given the double-peaked nature of the lines in Figs.~\ref{fig:varyrmin} and~\ref{fig:vars}, it seems plausible that two of the peaks in e.g. PDS 456's spectrum could come from a single absorber.
Reducing the required number of absorbers will be possible if a given absorber's blue and red peaks are not separated by more than the separation between a source's maximum and minimum absorber.
For example, the highest blueshifted absorber in PDS 456 has $g_{\rm max}=1.41$, whereas the least blueshifted absorber has $g_{\rm min}=1.26$ (Table~\ref{tab:absorbers}).
The {\it largest} ratio between the blue and red peaks that could condense the number of absorbers required by PDS 456 is $g_{\rm max}/g_{\rm min}=1.12$, shown by the gray shading in Fig.~\ref{fig:gratio}.
Only the lowest inclination angle $i=10^\circ$ can reach low enough energy ratios of the blue and red peak.
However, the energy shift of these peaks is not large enough to explain the energy shift required by the PDS 456 absorbers, shown by the blue shaded region.
Therefore, the minimally- and maximally-shifted absorbers in PDS 456 cannot be explained by a single ring absorber and we require five separate absorbers to explain the five absorption features.
For higher spin (not shown), the maximum $g$ extends beyond the shaded blue region (as demonstrated by Fig.~\ref{fig:maxg}) for high enough inclination angle, but the energy ratio remains well above the observed PDS 456 values, as anticipated from the right panel of Fig.~\ref{fig:vars}.

\subsection{Constraints on Absorber Width from Line Width}
To compare with observations of objects with absorption forests, we calculate the full width of the line feature at half of its maximum value (FWHM).
For strongly double-peaked line profiles (such as those at low inclination angles in Fig.~\ref{fig:vars} right panel), the FWHM does not carry a significant meaning.
As such, we calculate the FWHM of the line profiles shown in Fig.~\ref{fig:vars}'s central panel, which do have more clearly defined blue peaks.
To be consistent with XRISM observations, we calculate the FWHM for the energy range 6 - 10 keV. 

\begin{figure}
    \centering
    \includegraphics[width=0.9\linewidth]{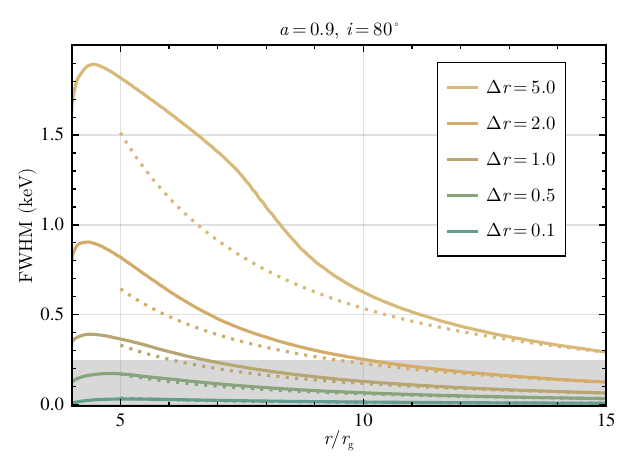}
    \caption{FWHM of the line between 6 - 10 keV, showing that only small absorber widths $\Delta r\lesssim 1r_g$ are consistent with the narrow widths of PDS 456. Dotted lines show the scaling expected from Keplerian velocity gradients $\Delta E\sim r^{-3/2}$, normalized to the FWHM values at $15r_g$. Gray region shows FWHM $<0.2$ keV, the approximate PDS 456 line widths. Converting the y-axis from energy shift to keV uses a rest-frame line energy of 6.67 keV (Fe XXV).}
    \label{fig:fwhm}
\end{figure}

Fig.~\ref{fig:fwhm} shows the dependence of the FWHM in terms of keV, using a rest-frame line energy of 6.67 keV as appropriate for Fe XXV.
As an absorbing ring with fixed width $\Delta r$ moves from larger radius $r_{\rm min}\approx 15r_g$ to $r_{\rm min}\approx 6r_g$, the line profile becomes broader mostly due to the velocity broadening introduced by sampling the Keplerian velocity at different radii.
The FWHM at $r_{\rm min}\gtrsim10r_g$ agrees well with the non-relativistic prediction $\Delta E\sim r^{-3/2}$, as shown by the dotted lines.
The approximation breaks down at smaller $r_{\rm min}$ as relativistic effects become important.
In particular, at $r\approx 4r_g$, Doppler beaming boosts the line profile's peak flux to an extremely narrow energy range, leading to the decrease in FWHM at the smallest radii shown. 
At $r_{\rm min}\lesssim 4r_g$, gravitational redshift and beaming combined produce line profiles that are not easily characterized by a FWHM. 
Because PDS 456 is best fit by absorbers at $r_{\rm min} \ge 5r_g$ (as demonstrated in Section~\ref{sec:pds456}), we exclude rings closer to the black hole from this plot. 
The FWHM's usefulness as a line width measure breaks down in that inner region anyway.
The Keplerian approximation also breaks down for $\Delta r=5r_g$, since the requirement that $\Delta r\ll r_{\rm min}$ is violated. 

Observationally, the absorption lines in PDS 456's X-ray transmission spectrum appear to have widths on the order of 0.2 keV or smaller (see Fig. 2 in~\citet{pds456}).
Fig.~\ref{fig:fwhm} shows that for $r_{\rm min}\gtrsim10r_g$, absorbing rings with $\Delta r\gtrsim2r_g$ produce lines that are broader than the observed range.
At smaller radii, the absorbing ring must have an even smaller width $\Delta r\lesssim0.5r_g$ to remain consistent with the PDS 456 widths.

Because, as discussed in the next section, the PDS 456 zones likely come from within $5-15r_g$ (Table~\ref{tab:absorbers}), the five absorbers must cover a region of $10r_g$, automatically constraining their widths as $\lesssim2r_g$.
Fig.~\ref{fig:fwhm} shows that the rings must be approximately a factor of 2 thinner than this estimate.

\section{Application to PDS 456} \label{sec:pds456}
Having outlined the characteristics of the line emission, we can construct a simple model for the forest of absorption lines seen by XRISM in PDS 456, a nearby ($z=0.184$), exceptionally bright radio-quiet quasar with bolometric luminosity on the order of $10^{47}$ erg/s~\citep{torres1997, reeves2000}.
In PDS 456, the five spectral features have rest-frame outflow velocities fit by~\citealt{pds456} given in Table~\ref{tab:absorbers}.
Because this velocity is inclination-angle-dependent, we convert back to energy shifts (Doppler factor) as $\mathcal{D}=\sqrt{1-\beta^2}/(1-\beta\cos\theta)$, where the inclination angle $\theta$ between the gas velocity and line of sight is fixed at $15^\circ$~\citep{pds456}.
We match each blueshift to an absorber with a fixed $\Delta r$ and a maximum energy shift $g$ as in Fig.~\ref{fig:maxg}, choosing $i=80^\circ$ and $a=0.9$ as an example.
For this set of parameters, the blueshifts come from absorbers with radii between 5 and 14$r_g$.

\begin{table}
\centering
\begin{tabular}{|l|c|c|c|c|}
\hline
 & $v_{\rm out}/c$ & $\mathcal{D}$($\theta=15^\circ$) & Absorber location $r/r_g$ & $t_{\rm orb}(r)/t_{\rm obs}$\\ \hline
Zone 1 & 0.226 & 1.246 & 13.5 & 2.30 \\ \hline
Zone 2 & 0.254 & 1.282 & 10.6 & 1.57\\ \hline
Zone 3 & 0.278 & 1.313 & 8.6 & 1.08 \\ \hline
Zone 4 & 0.307 & 1.353 & 6.8 & 0.67 \\ \hline
Zone 5 & 0.333 & 1.39 & 5.4 & 0.51\\ \hline
\end{tabular}
\caption{Summary of the PDS 456 absorption zone parameters. The observationally fit $v_{\rm out}/c$ (first column) from~\citet{pds456} are converted into Doppler factors (second column) using $\theta=15^\circ$ as stated in~\citet{pds456}. Those energy shifts are then matched to the total relativistic energy shifts from the location of an absorbing ring with $\Delta r=0.1r_g$ (column 3), assuming a spin of $a=0.9$ and inclination angle $i=80^\circ$; see Sec.~\ref{sec:pds456}. The right-most column gives the ratio of the orbital time assuming a black hole mass of $5\times10^8M_\odot$ to the XRISM observation duration of $t_{\rm obs}=250$ks.}
\label{tab:absorbers}
\end{table}
\begin{figure*}
    \centering
    \includegraphics[width=0.9\linewidth]{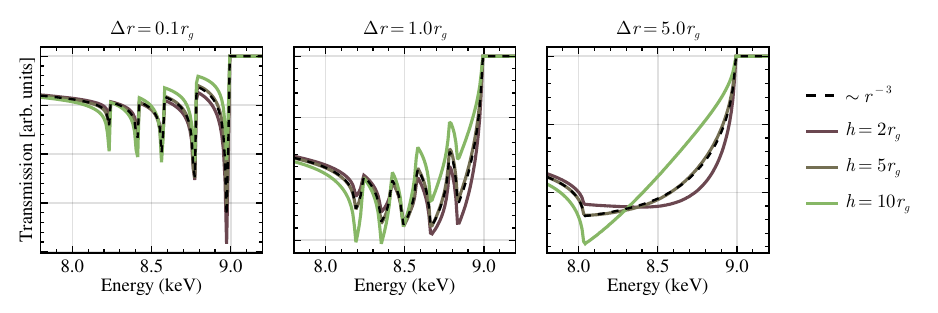}
    \caption{Example transmission spectra for five absorbers with the same blueshift as the PDS 456 features, for $a=0.9$ and $i=80^\circ$. Each panel assumes a fixed absorber width $\Delta r$ and a flat incoming spectrum. The x-axis is scaled to keV using a rest-frame line energy of 6.67 keV (Fe XXV). Solid lines show lamppost models, while the black dashed line shows $\sim r^{-3}$. The emissivity profiles of $r^{-3}$ and a lamppost with $h=5r_g$ are nearly identical.  \label{fig:transmission}}
\end{figure*}

Next, the line profiles from each annulus are combined into a spectrum using a specified emissivity profile.
In our calculations we use a ``lamppost'' corona~\citep{martocchia1996} emissivity profile with height $10r_g$ to match~\citet{chiang2017}, additional models with heights of $2r_g$ and $5r_g$, as well as the phenomenological $\propto r^{-3}$ emissivity profile~\citep{wilkins2012}.
The line profiles are then subtracted from a uniform background, which in principle includes the continuum emission and any broad emission lines, to construct the transmission spectrum shown in Fig.~\ref{fig:transmission}. 

We choose a highly spinning black hole with $a=0.9$ because for lower spins, the high energy shifts all come from close to the ISCO and thus blur together. 
The three panels of Fig.~\ref{fig:transmission} show the impact of different absorber widths, assuming that the absorption comes from a line with rest-frame energy of 6.67 keV as appropriate for Fe XXV. 
Values of $\Delta r\gtrsim 1.0 r_g$ (right panel) lose the clearly defined separation observed in e.g. PDS 456's spectrum, which comes from the separation of rings' blueshifted wings.
The dependence of the peak depth with energy depends sensitively on the emissivity, which determines the weighting of rings at different radii.
For an emissivity proportional to $r^{-3}$ (dashed black line), the rings at smaller radii dominate, leading to much deeper troughs at higher energies in narrow rings.
For wider rings, the wings of each line spread the troughs out over a larger range of energies (center panel).
For the lamppost emissivity, the height changes the relative depths of the absorption features because the emissivity profiles flatten at smaller radii, thereby increasing the relative depth of larger radii's blueshifted wings, even reversing the stepping pattern for a height of $10r_g$.  
A lamppost height of $5r_g$ is almost identical to an emissivity profile of $r^{-3}$ for the short radial range we consider.
The relative constancy of the trough depths in PDS 456 matches well with the emissivity profile of $r^{-3}$ and $\Delta r=1.0r_g$ (center panel dashed line).
This figure demonstrates that multiple absorption features are in principle possible.

The line-of-sight extent of the absorbing gas required to produce significant absorption lines can be estimated from parameters obtained by~\citet{pds456}.
We set the column density $N_H\approx 10^{23}~{\rm cm}^{-2}\approx n\Delta s$, where this value of $N_H$ does not include relativistic effects~\citep{pds456}.
Here, $n$ is the number density and $\Delta s$ is the path length through the absorbing cloud, i.e. roughly its extent along the line-of-sight.
Note that the optical depth to scattering $\tau=n\sigma_T\Delta s=N_H\sigma_T$, where $\sigma_T$ is the Thomson cross-section, has the same value of $\tau\approx0.1$ as the discovery paper, as is the line opacity, since the absorbers are assigned the column density and ionization parameter of that paper's photoionization fit~\citep{pds456}. 
In this sense, the disc-origin model changes only where the absorbing gas is located and how it moves rather than how much of it there is or how it absorbs.
Taking the same ionizing luminosity $L_X\approx10^{46}$ erg/s, the fit value $\xi=10^5$ erg cm/s~\citep{pds456}, and $R\approx10r_g$ from the absorber locations (Table~\ref{tab:absorbers}), the relation $\xi=L_X/(nR^2)=L_X\Delta s/(N_HR^2)$ gives
\begin{align}
    \frac{\Delta s}{5\times10^{11}~{\rm cm}}&=\left(\frac{L_{\rm ion}}{10^{46}~{\rm erg/s}}\right)^{-1}\left(\frac{\xi}{10^5~{\rm erg~cm/s}}\right)\left(\frac{N_H}{10^{23}~{\rm cm}^{-2}}\right)\left(\frac{R}{10r_g}\right)^2\label{eq:deltas}
\end{align}
where $r_g=7\times10^{13}~{\rm cm}$ for the assumed black hole mass of $5\times10^8M_\odot$. Eq.~\ref{eq:deltas} shows that the gas must have a line-of-sight (LoS) path length of roughly $\Delta s=7\times10^{-3}r_g$.

\subsection{Inclination Angle from the Broad Emission Feature}
In order to explain the blueshifted absorption features using this model of material lying just above the accretion disc, the disc must be observed close to edge-on, with the normal to the disc inclined at an angle greater than approximately $75^\circ$.

We find that the broad emission feature observed in the iron K band of the \textit{XRISM Resolve} spectrum of PDS\,456 can be explained as the relativistically broadened iron K$\alpha$ fluorescence line, produced in the reflection spectrum as the primary X-ray emission from the corona irradiates, and is reprocessed by, the inner regions of the accretion disc. The broadening of the line by the combination of Doppler shifts and gravitational redshift means that the blueshifted edge of the line provides constraints on the inclination of the accretion disc. Similarly, \citet{chiang2017} find that the \textit{Suzaku} spectrum of PDS\,456 can be fit with a similar relativistic reflection model. Most significantly, they detect a time lag between correlated variations in the X-ray continuum and iron K band of $\sim 10^4$\,s. Assuming a black hole mass of $8\times 10^8\,M_\odot$~\citep{nardini2015, li2024}, this lag is consistent with X-ray reverberation from the inner accretion disc, suggesting that the broad iron K emission feature is produced within 3\,$r_\mathrm{g}$ of the black hole, in line with the reflection model.

To estimate the range of disc inclinations that are consistent with the observed spectrum, we fit a \textsc{relxilllpCp} model~\citep{dauser+2014,garcia+2014} spectrum in which the continuum and broad emission feature are explained by the X-ray emission from a compact corona (modeled as a simplified point source located on the spin axis above the black hole) and the relativistic reflection of this continuum from the inner accretion disc. 
In this section, we do not attempt to apply the full model of absorption arising from the accretion disc. Rather, we seek to demonstrate that the broad emission feature in the X-ray spectrum of PDS\,456 is consistent with the high inclination angles required to produce the required line-of-sight velocities in clouds tied to the disc. We therefore employ a phenomenological description of the absorption line forest consisting of five photoionized absorbers modeled using \textsc{xstar}, for each of which the velocities and column densities are fit freely to the observed spectrum. 
As in \citet{pds456}, the components are assumed to have the same ionization parameter, which is tied between the components but still fit freely to the data. 
We thus marginalize over any uncertainties in the absorption model. 
Details of the data reduction and modeling can be found in Appendix~\ref{app:modeling}.

\begin{figure}
    \centering
    \includegraphics[width=0.9\linewidth]{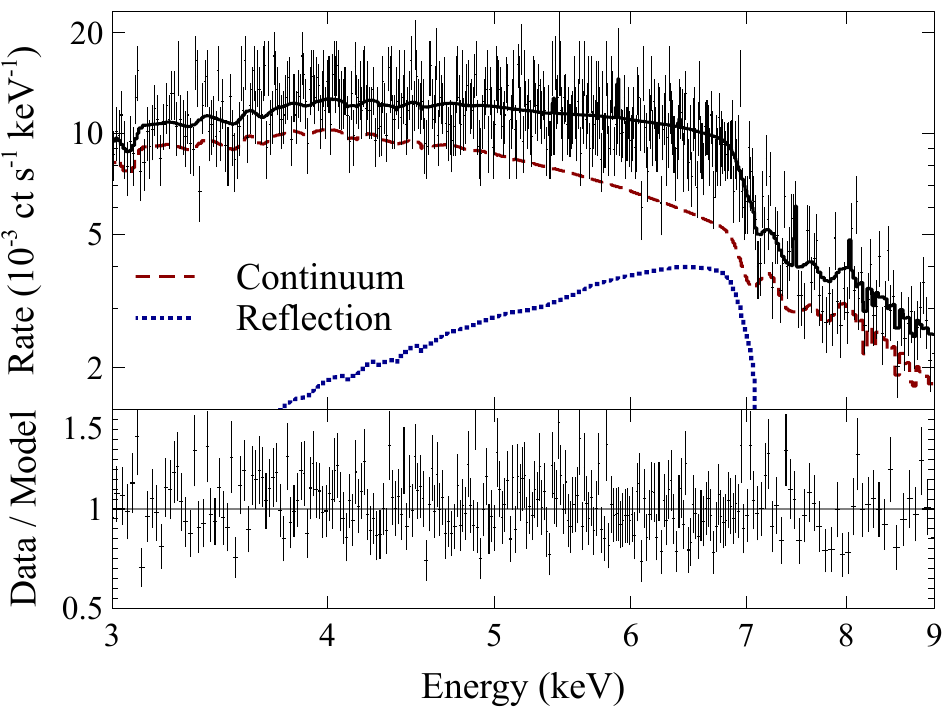}
\caption{The \textit{XRISM Resolve} spectrum of PDS\,456 fit with a model in which the broad emission component is produced by reflection from the inner accretion disc, described using the \textsc{relxilllpCp} model. The absorption lines are modeled by five photoionized components using \textsc{xstar}. The best-fitting inclination of the accretion disc is $69_{-2}^{+7}$\,deg. The model provides a good fit to the data, yielding a $C$-statistic of 2476 for 2372 degrees of freedom.}
    \label{fig:relxill_fit}
\end{figure}

This model, in which the broad emission component is produced by reflection from the inner accretion disc, provides an acceptable fit to the \textit{XRISM Resolve} spectrum in the 2-12\,keV band (Fig.~\ref{fig:relxill_fit}), yielding a $C$-statistic of 2476 for 2372 degrees of freedom. Under the simplifying assumptions of this reflection model, the best-fitting inclination of the accretion disc is $69_{-2}^{+7}$\,deg (where the uncertainty corresponds to the 90 per cent confidence interval), roughly consistent with the inclination angle constraint required for the disc absorption model (Fig.~\ref{fig:incconstraint}). 

We use the point source approximation to provide a phenomenological description of the accretion disc emissivity profile, applicable when the corona is compact with respect to the scale of the reflecting accretion disc \citep[e.g.][]{wilkins2012}. If, instead, we fit the accretion disc emissivity profile as a broken power law to allow for a generalized coronal geometry, we obtain a marginal improvement in the fit, with $\Delta C = -9$ for 2 additional free parameters. The best-fitting power law emissivity profile is steep, falling as $r^{-6}$ within the inner 3\,$r_\mathrm{g}$, and $r^{-3.5}$ over the outer disc, showing that this model is consistent with the corona being compact. The constraint on the inclination of the accretion disc is consistent between the two models. With the broken power law emissivity model, we find the inclination to be $69_{-4}^{+4}$\,deg. Detailed modeling of the nature of the primary X-ray source is beyond the scope of this work.

\section{Discussion}
\subsection{Origin of the Absorbing Clouds} \label{ssec:origin}
Although fully detailing the origin of the dense absorbing clouds is beyond the scope of this work, we envision a physically motivated, qualitative scenario.
In this scenario, magnetic flux tubes from the disc midplane buoyantly rise into the disc atmosphere via the Parker instability~\citep{parker1966}.
These flux tubes drag dense gas upward, suspending it until it shears out on an orbital timescale or dissipates through magnetic reconnection.
This scenario draws an analogy to solar prominences, where an uplifted magnetic flux bundle can twist, buckle, and hold a packet of dense gas in place for extended timescales~\citep{zhou2025}.
The same tension that confines the gas against gravity also resists radiation pressure, such that the clouds need not be expelled at near-Eddington luminosity.
The observed increase of absorption line velocity with X-ray luminosity~\citep{matzeu2017} requires that the absorbing radius moves inward as the source brightens.

Although we have assumed an accretion disc that is infinitely thin in the vertical direction~\citep{ss73}, PDS 456's luminosity suggests a super-Eddington accretion rate~\citep{nardini2015}.
Simulations of the super-Eddington regime have found atmospheres that fragment into clumps~\citep{takeuchi2013}, forming dense condensations in an otherwise hot flow~\citep{liska2022, bambic2024} that could provide the absorbing gas our model requires.

In principle, the absorbers could arise from localized column density enhancements, such as those in gravitationally-bound failed winds.
However, the even frequency spacing of the absorption signatures points towards overdensities driven by disc instabilities, which are subsequently picked out by relativistic beaming along the line-of-sight.
Future work will investigate the presence of these loops in general relativistic magnetohydrodynamic simulations and implications for coronal heating.

\subsection{Absorber Shape and Radial Extent} \label{ssec:shape}
Although we have considered absorbing rings for computational ease, the azimuthal shape of the absorber does not necessarily determine the resulting absorption signatures. 
As shown in Fig.~\ref{fig:schematic}'s right panel, only a small azimuthal portion of a ring at a given radius contributes the necessary blueshifts.
As~\citet{fabian2020} pointed out, the brightest part of the reflection spectrum comes from slightly behind the black hole for higher inclination angles.
In this case, the absorbers could constitute small blobs of dense gas that happen to travel towards the observer during an observation time.
The right-most column in Table~\ref{tab:absorbers} shows that the three innermost absorbers make at least one complete orbit during the XRISM observation of PDS 456, increasing the likelihood of creating an observable blueshifted absorption zone.

\subsection{High Inclination Angle Constraints}
The absorbing disc model requires fairly strict constraints on inclination angle in order to produce energy shifts consistent with PDS 456 (Fig.~\ref{fig:incconstraint}, Table~\ref{tab:absorbers}).
%
The strict inclination angle requirements could be slightly relaxed if the absorption comes from  Fe XXVI rather than Fe XXV. 
In this case, the highest energy line at 9.3 keV would have an energy shift of $g=1.33$ rather than 1.39, which would require inclination angles $i\gtrsim70^\circ$ depending on spin (Fig.~\ref{fig:incconstraint}).
Determining which line dominates the absorption spectrum requires full photoionization modeling of the absorbing cloud system, which is beyond the scope of this paper.
Assuming that Fe XXV dominates the absorption therefore provides upper limits on the inclination angle.
Notably, the presence of both Fe XXV and Fe XXVI could reduce the number of absorbing zones required to produce the five distinct lines; again, neglecting Fe XXVI therefore provides an upper limit.

Super-Eddington effects could also affect inclination angle measurements.
Finite disc height could lead to truncation of the lines' blue wings~\citep{taylor2018a, gradus} and affect the line lags~\citep{taylor2018b}.
If the accretion disc is more akin to a puffy, slim disc model~\citep{abramowicz1988}, lower inclination angles might be required to minimize obscuration effects that will lower the maximum blueshift of the line~\citep{shashank2026}.
A transition between a thin disc and a hot inner flow could happen within $5r_g$~\citep{hankla2025}, potentially providing a source of continuum photons without affecting the absorber location.
Any change in the disc or corona geometry, such as a tilting of the inner disc, could also affect the inclination angle constraints.

\subsection{Absorber Variability}
Variability in the absorption spectrum could be due to both formation and destruction of the dense clumps of gas.
If the absorbers occur due to buoyancy of magnetic flux tubes dragging dense gas upwards as proposed in Sec.~\ref{ssec:origin}, their emergence timescale is likely dictated by the Parker instability's growth rate.
The instability's growth rate will depend on the local disc conditions, especially the ratio of thermal gas pressure to magnetic pressure. 
Because this growth rate can be faster than an orbital time $t_{\rm orb}\equiv2\pi/\Omega_K$, where $\Omega_K\sim r^{-3/2}$ is the Keplerian angular frequency, these absorption features can appear within a XRISM observation.

Once created, dense gas tied to the accretion disc's orbital motion might break up or viscously spread over a finite timescale. 
Assuming a thin accretion disc~\citep{ss73}, the viscous timescale for perturbations to propagate to smaller radii is a factor of $\alpha^{-1}(H/r)^{-2}$ times the orbital timescale.
For modest values of effective viscosity $\alpha\approx0.1$ and a disc scaleheight ratio $H/r\approx$0.01, the viscous timescale is $10^5$ times the orbital timescale, and therefore will likely not significantly affect the radial extent of the absorbers over the course of a XRISM observation. 

Although the radial spreading of the dense gas is likely inefficient, an initially localized gas packet could become sheared out azimuthally on timescales close to the orbital time.
In such a scenario, the absorbing zones could disappear and reappear on the orbital timescale, producing variability on the order of hundreds of kiloseconds, comparable to XRISM observation durations (Table~\ref{tab:absorbers}). 
Indeed, the XRISM observation of PDS 456 shows the absorbing zones moving from discrete zones to a continuous spread over the course of 100 - 200 ks (not shown).
Variability on the timescale of days has also been seen in PG 1211~\citep{pounds2025}, another quasar with multiple absorption zones~\citep{mizumoto2026, reeves2026}, and was suggested to be the result of rapidly accreting (and thus redshifting) material~\citep{pounds2024}.
Upcoming work will explore using this variability to discriminate between the UFO and disc-absorption models.
Simulations of accretion discs have already suggested non-axisymmetric structures as mechanisms for producing variable amounts of light~\citep{armitage2003}.

If the absorbers move around the disc with the Keplerian orbital velocity, they could also produce time-varying absorption lines at $\approx 4$ keV due to the redshifted peaks of the blurred lines. 
Although hints of absorption lines at 4 keV are present in the XRISM data, more work is needed to determine the significance of their detection.
Other multiple blue-and red-shifted absorption lines such as those found in Mrk 509~\citep{dadina2026} could be consistent with our model.

\section{Conclusions}
Exploring alternatives to ultrafast outflows serves a crucial role in testing and validating the impact of these outflows on feedback and galaxy evolution.
Here, we have proposed one such alternative model, wherein the relativistic bulk motion of the absorber originates in the accretion disc itself rather than an outflow in the polar regions of the system.
Clumps of dense, absorbing gas originate from disc instabilities, perhaps tied to buoyantly rising magnetic flux tubes.
A first exploration of the disc motion model with five narrow ($\lesssim1r_g$) clumps of gas within $\approx 15r_g$ of the central black hole can explain the separated troughs in PDS 456's transmission spectrum given inclination angles $i\gtrsim75^\circ$.
This model therefore merits further investigation, in particular into the nature of these absorbers and the plausibility of their formation in an accretion disc. 
A deeper exploration of absorption forests in PDS 456 and similar objects could reveal the detailed innermost structure and behaviour of luminous accreting supermassive black holes.

\section*{Acknowledgements}
The authors thank C. Reynolds for helpful discussions.
Support for this work was provided by NASA through the NASA Hubble Fellowship grant \#HF2-51555 awarded by the Space Telescope Science Institute, which is operated by the Association of Universities for Research in Astronomy, Inc., for NASA, under contract NAS5-26555. 
FJEB acknowledges support by the European Union (ERC, X-MAPS, 101169908). 
AMH acknowledges the use of Copilot in VSCode to correct Julia language typos and ChatGPT to identify typographic errors in the final stages of revision.

\section*{Data Availability}
The data in this article were produced using the open-access code~\href{https://codeberg.org/astro-group/Gradus.jl}{\texttt{GRADUS.JL}}~\citep{gradus}. 
The full dataset is available upon reasonable request to the authors.



\bibliographystyle{mnras}
\bibliography{refs} 

@article{garcia+2014,
	adsurl = {https://ui.adsabs.harvard.edu/abs/2014ApJ...782...76G},
	archiveprefix = {arXiv},
	author = {{Garc{\'\i}a}, J. and {Dauser}, T. and {Lohfink}, A. and {Kallman}, T.~R. and {Steiner}, J.~F. and {McClintock}, J.~E. and {Brenneman}, L. and {Wilms}, J. and {Eikmann}, W. and {Reynolds}, C.~S. and {Tombesi}, F.},
	doi = {10.1088/0004-637X/782/2/76},
	eid = {76},
	eprint = {1312.3231},
	journal = {\apj},
	month = {Feb},
	number = {2},
	pages = {76},
	primaryclass = {astro-ph.HE},
	title = {{Improved Reflection Models of Black Hole Accretion Disks: Treating the Angular Distribution of X-Rays}},
	volume = {782},
	year = {2014}
}

@ARTICLE{Travascio2024,
       author = {{Travascio}, A. and {Piconcelli}, E. and {Bischetti}, M. and {Cresci}, G. and {Feruglio}, C. and {Perna}, M. and {Vietri}, G. and {Carniani}, S. and {Cantalupo}, S. and {Cicone}, C. and {Ginolfi}, M. and {Venturi}, G. and {Zubovas}, K. and {Bongiorno}, A. and {Brusa}, M. and {Luminari}, A. and {Mainieri}, V. and {Marconi}, A. and {Menci}, N. and {Nardini}, E. and {Pensabene}, A. and {Ramos Almeida}, C. and {Tombesi}, F. and {Vignali}, C. and {Zappacosta}, L. and {Fiore}, F.},
        title = "{MUSE view of PDS 456: Kiloparsec-scale wind, extended ionized gas, and close environment}",
      journal = {\aap},
         year = 2024,
        month = jun,
       volume = {686},
          eid = {A250},
        pages = {A250},
          doi = {10.1051/0004-6361/202449201},
archivePrefix = {arXiv},
       eprint = {2403.18043},
 primaryClass = {astro-ph.GA},
       adsurl = {https://ui.adsabs.harvard.edu/abs/2024A&A...686A.250T}
}

@ARTICLE{Takeuchi2013,
       author = {{Takeuchi}, Shun and {Ohsuga}, Ken and {Mineshige}, Shin},
        title = "{Clumpy Outflows from Supercritical Accretion Flow}",
      journal = {\pasj},
         year = 2013,
        month = aug,
       volume = {65},
          eid = {88},
        pages = {88},
          doi = {10.1093/pasj/65.4.88},
archivePrefix = {arXiv},
       eprint = {1305.1023},
 primaryClass = {astro-ph.HE},
       adsurl = {https://ui.adsabs.harvard.edu/abs/2013PASJ...65...88T}
}

@ARTICLE{Bambic2024,
       author = {{Bambic}, Christopher J. and {Quataert}, Eliot and {Kunz}, Matthew W.},
        title = "{Local models of two-temperature accretion disc coronae - I. Structure, outflows, and energetics}",
      journal = {\mnras},
         year = 2024,
        month = jan,
       volume = {527},
       number = {2},
        pages = {2895-2918},
          doi = {10.1093/mnras/stad3261},
archivePrefix = {arXiv},
       eprint = {2304.06067},
 primaryClass = {astro-ph.HE},
       adsurl = {https://ui.adsabs.harvard.edu/abs/2024MNRAS.527.2895B}
}

@ARTICLE{Matzeu2017,
       author = {{Matzeu}, G.~A. and {Reeves}, J.~N. and {Braito}, V. and {Nardini}, E. and {McLaughlin}, D.~E. and {Lobban}, A.~P. and {Tombesi}, F. and {Costa}, M.~T.},
        title = "{Evidence for a radiatively driven disc-wind in PDS 456?}",
      journal = {\mnras},
         year = 2017,
        month = nov,
       volume = {472},
       number = {1},
        pages = {L15-L19},
          doi = {10.1093/mnrasl/slx129},
archivePrefix = {arXiv},
       eprint = {1708.03546},
 primaryClass = {astro-ph.HE},
       adsurl = {https://ui.adsabs.harvard.edu/abs/2017MNRAS.472L..15M}
}

@ARTICLE{Liska2022,
       author = {{Liska}, M.~T.~P. and {Musoke}, G. and {Tchekhovskoy}, A. and {Porth}, O. and {Beloborodov}, A.~M.},
        title = "{Formation of Magnetically Truncated Accretion Disks in 3D Radiation-transport Two-temperature GRMHD Simulations}",
      journal = {\apjl},
         year = 2022,
        month = aug,
       volume = {935},
       number = {1},
          eid = {L1},
        pages = {L1},
          doi = {10.3847/2041-8213/ac84db},
archivePrefix = {arXiv},
       eprint = {2201.03526},
 primaryClass = {astro-ph.HE},
       adsurl = {https://ui.adsabs.harvard.edu/abs/2022ApJ...935L...1L}
}

@ARTICLE{Seebeck2024,
       author = {{Seebeck}, Jerome and {Veilleux}, Sylvain and {Liu}, Weizhe and {Rupke}, David S.~N. and {Vayner}, Andrey and {Wylezalek}, Dominika and {Zakamska}, Nadia L. and {Bertemes}, Caroline},
        title = "{Combined JWST─MUSE Integral Field Spectroscopy of the Most Luminous Quasar in the Local Universe, PDS 456}",
      journal = {\apj},
         year = 2024,
        month = dec,
       volume = {976},
       number = {2},
          eid = {240},
        pages = {240},
          doi = {10.3847/1538-4357/ad815e},
archivePrefix = {arXiv},
       eprint = {2409.18086},
 primaryClass = {astro-ph.GA},
       adsurl = {https://ui.adsabs.harvard.edu/abs/2024ApJ...976..240S}
}

@ARTICLE{Bischetti2019,
       author = {{Bischetti}, M. and {Piconcelli}, E. and {Feruglio}, C. and {Fiore}, F. and {Carniani}, S. and {Brusa}, M. and {Cicone}, C. and {Vignali}, C. and {Bongiorno}, A. and {Cresci}, G. and {Mainieri}, V. and {Maiolino}, R. and {Marconi}, A. and {Nardini}, E. and {Zappacosta}, L.},
        title = "{The gentle monster PDS 456. Kiloparsec-scale molecular outflow and its implications for QSO feedback}",
      journal = {\aap},
         year = 2019,
        month = aug,
       volume = {628},
          eid = {A118},
        pages = {A118},
          doi = {10.1051/0004-6361/201935524},
archivePrefix = {arXiv},
       eprint = {1903.10528},
 primaryClass = {astro-ph.GA},
       adsurl = {https://ui.adsabs.harvard.edu/abs/2019A&A...628A.118B}
}

@ARTICLE{Reeves2018,
       author = {{Reeves}, J.~N. and {Braito}, V. and {Nardini}, E. and {Lobban}, A.~P. and {Matzeu}, G.~A. and {Costa}, M.~T.},
        title = "{A New Relativistic Component of the Accretion Disk Wind in PDS 456}",
      journal = {\apjl},
         year = 2018,
        month = feb,
       volume = {854},
       number = {1},
          eid = {L8},
        pages = {L8},
          doi = {10.3847/2041-8213/aaaae1},
archivePrefix = {arXiv},
       eprint = {1801.08899},
 primaryClass = {astro-ph.HE},
       adsurl = {https://ui.adsabs.harvard.edu/abs/2018ApJ...854L...8R}
}

@ARTICLE{boissay2019,
       author = {{Boissay-Malaquin}, Rozenn and {Danehkar}, Ashkbiz and {Marshall}, Herman L. and {Nowak}, Michael A.},
        title = "{Relativistic Components of the Ultra-fast Outflow in the Quasar PDS 456 from Chandra/HETGS, NuSTAR, and XMM-Newton Observations}",
      journal = {\apj},
         year = 2019,
        month = mar,
       volume = {873},
       number = {1},
          eid = {29},
        pages = {29},
          doi = {10.3847/1538-4357/ab0082},
archivePrefix = {arXiv},
       eprint = {1901.06641},
 primaryClass = {astro-ph.HE},
       adsurl = {https://ui.adsabs.harvard.edu/abs/2019ApJ...873...29B}
}

@ARTICLE{dauser+2014,
       author = {{Dauser}, T. and {Garcia}, J. and {Parker}, M.~L. and {Fabian}, A.~C. and {Wilms}, J.},
        title = "{The role of the reflection fraction in constraining black hole spin.}",
      journal = {\mnras},
         year = 2014,
        month = oct,
       volume = {444},
        pages = {L100-L104},
          doi = {10.1093/mnrasl/slu125},
archivePrefix = {arXiv},
       eprint = {1408.2347},
 primaryClass = {astro-ph.HE},
       adsurl = {https://ui.adsabs.harvard.edu/abs/2014MNRAS.444L.100D}
}

@ARTICLE{hameury1994,
       author = {{Hameury}, J. -M. and {Marck}, J. -A. and {Pelat}, D.},
        title = "{E\^+\^ - e\^-\^ annihilation lines from accretion discs around Kerr black holes}",
      journal = {\aap},
         year = 1994,
        month = jul,
       volume = {287},
        pages = {795-802},
       adsurl = {https://ui.adsabs.harvard.edu/abs/1994A&A...287..795H}
}

@ARTICLE{gates2025,
       author = {{Gates}, Delilah E.~A. and {Truong}, Chau and {Sahu}, Amrita and {C{\'a}rdenas-Avenda{\~n}o}, Alejandro},
        title = "{Morphology of relativistically broadened line emission from axisymmetric equatorial accretion disks}",
      journal = {\prd},
         year = 2025,
        month = jun,
       volume = {111},
       number = {12},
          eid = {124004},
        pages = {124004},
          doi = {10.1103/PhysRevD.111.124004},
archivePrefix = {arXiv},
       eprint = {2411.14338},
 primaryClass = {astro-ph.HE},
       adsurl = {https://ui.adsabs.harvard.edu/abs/2025PhRvD.111l4004G}
}

@ARTICLE{gallo2013,
       author = {{Gallo}, L.~C. and {Fabian}, A.~C.},
        title = "{The origin of blueshifted absorption features in the X-ray spectrum of PG  1211+143: outflow or disc.}",
      journal = {\mnras},
         year = 2013,
        month = jul,
       volume = {434},
        pages = {L66-L69},
          doi = {10.1093/mnrasl/slt080},
archivePrefix = {arXiv},
       eprint = {1306.3404},
 primaryClass = {astro-ph.HE},
       adsurl = {https://ui.adsabs.harvard.edu/abs/2013MNRAS.434L..66G}
}

@ARTICLE{gallo2011,
       author = {{Gallo}, L.~C. and {Fabian}, A.~C.},
        title = "{How the effects of resonant absorption on black hole reflection spectra can mimic high-velocity outflows}",
      journal = {\mnras},
         year = 2011,
        month = nov,
       volume = {418},
       number = {1},
        pages = {L59-L63},
          doi = {10.1111/j.1745-3933.2011.01143.x},
archivePrefix = {arXiv},
       eprint = {1108.5060},
 primaryClass = {astro-ph.HE},
       adsurl = {https://ui.adsabs.harvard.edu/abs/2011MNRAS.418L..59G}
}

@ARTICLE{reeves2026,
       author = {{Reeves}, James and {Braito}, Valentina and {Mizumoto}, Misaki and {Kraemer}, Steven and {Behar}, Ehud and {Done}, Chris and {Hagino}, Kouichi and {Matzeu}, Gabriele and {Noda}, Hirofumi and {Nomura}, Mariko and {Ogawa}, Shoji and {Ohsuga}, Ken and {Tanimoto}, Atsushi and {Turner}, Tracey and {Ueda}, Yoshihiro and {Yamada}, Satoshi and {Ganguly}, Sreeparna and {Somenzi}, Paolo and {Reich}, Omer},
        title = "{Resolving the Multiple Component Outflows in PG 1211+143: II. The Soft X-ray View of the Ultra Fast Outflow}",
      journal = {arXiv e-prints},
         year = 2026,
        month = feb,
          eid = {arXiv:2602.16496},
        pages = {arXiv:2602.16496},
          doi = {10.48550/arXiv.2602.16496},
archivePrefix = {arXiv},
       eprint = {2602.16496},
 primaryClass = {astro-ph.HE},
       adsurl = {https://ui.adsabs.harvard.edu/abs/2026arXiv260216496R}
}

@ARTICLE{shashank2026,
       author = {{Shashank}, Swarnim and {Abdikamalov}, Askar B. and {Liu}, Honghui and {Mirzaev}, Temurbek and {Jiang}, Jiachen and {Bambi}, Cosimo and {Baker}, Fergus and {Young}, Andrew},
        title = "{Modeling Reflection Spectra of Super-Eddington X-Ray Sources}",
      journal = {\apj},
         year = 2026,
        month = jan,
       volume = {996},
       number = {1},
          eid = {51},
        pages = {51},
          doi = {10.3847/1538-4357/ae2614},
archivePrefix = {arXiv},
       eprint = {2407.12890},
 primaryClass = {astro-ph.HE},
       adsurl = {https://ui.adsabs.harvard.edu/abs/2026ApJ...996...51S}
}

@ARTICLE{gradus,
       author = {{Baker}, F.~J.~E. and {Young}, A.~J.},
        title = "{GRADUS.JL: spacetime-agnostic general relativistic ray-tracing for X-ray spectral modelling}",
      journal = {\mnras},
         year = 2026,
        month = jan,
       volume = {545},
       number = {2},
          eid = {staf1770},
        pages = {staf1770},
          doi = {10.1093/mnras/staf1770},
archivePrefix = {arXiv},
       eprint = {2510.15049},
 primaryClass = {astro-ph.HE},
       adsurl = {https://ui.adsabs.harvard.edu/abs/2026MNRAS.545f1770B}
}

@ARTICLE{xrism,
       author = {{XRISM Science Team}},
        title = "{Science with the X-ray Imaging and Spectroscopy Mission (XRISM)}",
      journal = {arXiv e-prints},
         year = 2020,
        month = mar,
          eid = {arXiv:2003.04962},
        pages = {arXiv:2003.04962},
          doi = {10.48550/arXiv.2003.04962},
archivePrefix = {arXiv},
       eprint = {2003.04962},
 primaryClass = {astro-ph.HE},
       adsurl = {https://ui.adsabs.harvard.edu/abs/2020arXiv200304962X}
}

@ARTICLE{mizumoto2026,
       author = {{Mizumoto}, Misaki and {Reeves}, James N. and {Braito}, Valentina and {Behar}, Ehud and {Done}, Chris and {Hagino}, Kouichi and {Kraemer}, Steven B. and {Matzeu}, Gabriele A. and {Noda}, Hirofumi and {Nomura}, Mariko and {Ogawa}, Shoji and {Ohsuga}, Ken and {Tanimoto}, Atsushi and {Turner}, Tracey J. and {Ueda}, Yoshihiro and {Yamada}, Satoshi and {Ganguly}, Sreeparna and {Somenzi}, Paolo},
        title = "{Resolving the Multiple Component Outflows in PG 1211+143. I. The Fe─K Absorption Structure and UFO Forest}",
      journal = {\apj},
         year = 2026,
        month = feb,
       volume = {997},
       number = {2},
          eid = {219},
        pages = {219},
          doi = {10.3847/1538-4357/ae2853},
archivePrefix = {arXiv},
       eprint = {2512.03533},
 primaryClass = {astro-ph.HE},
       adsurl = {https://ui.adsabs.harvard.edu/abs/2026ApJ...997..219M}
}

@ARTICLE{pounds2024,
       author = {{Pounds}, Ken and {Page}, Kim},
        title = "{Low-redshift absorption in the Seyfert galaxy PG1211+143 - a distant inflow maintaining off-plane accretion or the gravitational redshift of matter orbiting the SMBH?}",
      journal = {\mnras},
         year = 2024,
        month = jul,
       volume = {531},
       number = {4},
        pages = {4852-4856},
          doi = {10.1093/mnras/stae1491},
archivePrefix = {arXiv},
       eprint = {2311.09853},
 primaryClass = {astro-ph.HE},
       adsurl = {https://ui.adsabs.harvard.edu/abs/2024MNRAS.531.4852P}
}

@ARTICLE{ss73,
       author = {{Shakura}, N.~I. and {Sunyaev}, R.~A.},
        title = "{Black holes in binary systems. Observational appearance.}",
      journal = {\aap},
         year = 1973,
        month = jan,
       volume = {24},
        pages = {337-355},
       adsurl = {https://ui.adsabs.harvard.edu/abs/1973A&A....24..337S}
}

@ARTICLE{taylor2018b,
       author = {{Taylor}, Corbin and {Reynolds}, Christopher S.},
        title = "{X-Ray Reverberation from Black Hole Accretion Disks with Realistic Geometric Thickness}",
      journal = {\apj},
         year = 2018,
        month = dec,
       volume = {868},
       number = {2},
          eid = {109},
        pages = {109},
          doi = {10.3847/1538-4357/aae9f2},
archivePrefix = {arXiv},
       eprint = {1808.07877},
 primaryClass = {astro-ph.HE},
       adsurl = {https://ui.adsabs.harvard.edu/abs/2018ApJ...868..109T}
}

@ARTICLE{taylor2018a,
       author = {{Taylor}, Corbin and {Reynolds}, Christopher S.},
        title = "{Exploring the Effects of Disk Thickness on the Black Hole Reflection Spectrum}",
      journal = {\apj},
         year = 2018,
        month = mar,
       volume = {855},
       number = {2},
          eid = {120},
        pages = {120},
          doi = {10.3847/1538-4357/aaad63},
archivePrefix = {arXiv},
       eprint = {1712.05418},
 primaryClass = {astro-ph.HE},
       adsurl = {https://ui.adsabs.harvard.edu/abs/2018ApJ...855..120T}
}

@ARTICLE{wilkins2012,
       author = {{Wilkins}, D.~R. and {Fabian}, A.~C.},
        title = "{Understanding X-ray reflection emissivity profiles in AGN: locating the X-ray source}",
      journal = {\mnras},
         year = 2012,
        month = aug,
       volume = {424},
       number = {2},
        pages = {1284-1296},
          doi = {10.1111/j.1365-2966.2012.21308.x},
archivePrefix = {arXiv},
       eprint = {1205.3179},
 primaryClass = {astro-ph.HE},
       adsurl = {https://ui.adsabs.harvard.edu/abs/2012MNRAS.424.1284W}
}

@ARTICLE{martocchia1996,
       author = {{Martocchia}, Andrea and {Matt}, Giorgio},
        title = "{Iron Kalpha line intensity from accretion discs around rotating black holes}",
      journal = {\mnras},
         year = 1996,
        month = oct,
       volume = {282},
       number = {4},
        pages = {L53-L57},
          doi = {10.1093/mnras/282.4.L53},
       adsurl = {https://ui.adsabs.harvard.edu/abs/1996MNRAS.282L..53M}
}

@ARTICLE{hankla2025,
       author = {{Hankla}, Amelia M. and {Dexter}, Jason and {Scepi}, Nicolas},
        title = "{The inner structure and thermodynamics of thin accretion discs}",
      journal = {\mnras},
         year = 2025,
        month = aug,
       volume = {541},
       number = {4},
        pages = {3184-3197},
          doi = {10.1093/mnras/staf1169},
archivePrefix = {arXiv},
       eprint = {2504.21207},
 primaryClass = {astro-ph.HE},
       adsurl = {https://ui.adsabs.harvard.edu/abs/2025MNRAS.541.3184H}
}

@ARTICLE{armitage2003,
       author = {{Armitage}, Philip J. and {Reynolds}, Christopher S.},
        title = "{The variability of accretion on to Schwarzschild black holes from turbulent magnetized discs}",
      journal = {\mnras},
         year = 2003,
        month = may,
       volume = {341},
       number = {3},
        pages = {1041-1050},
          doi = {10.1046/j.1365-8711.2003.06491.x},
archivePrefix = {arXiv},
       eprint = {astro-ph/0302271},
 primaryClass = {astro-ph},
       adsurl = {https://ui.adsabs.harvard.edu/abs/2003MNRAS.341.1041A}
}

@ARTICLE{nardini2015,
       author = {{Nardini}, E. and {Reeves}, J.~N. and {Gofford}, J. and {Harrison}, F.~A. and {Risaliti}, G. and {Braito}, V. and {Costa}, M.~T. and {Matzeu}, G.~A. and {Walton}, D.~J. and {Behar}, E. and {Boggs}, S.~E. and {Christensen}, F.~E. and {Craig}, W.~W. and {Hailey}, C.~J. and {Matt}, G. and {Miller}, J.~M. and {O'Brien}, P.~T. and {Stern}, D. and {Turner}, T.~J. and {Ward}, M.~J.},
        title = "{Black hole feedback in the luminous quasar PDS 456}",
      journal = {Science},
         year = 2015,
        month = feb,
       volume = {347},
       number = {6224},
        pages = {860-863},
          doi = {10.1126/science.1259202},
archivePrefix = {arXiv},
       eprint = {1502.06636},
 primaryClass = {astro-ph.HE},
       adsurl = {https://ui.adsabs.harvard.edu/abs/2015Sci...347..860N}
}

@ARTICLE{zhou2025,
       author = {{Zhou}, Yuhao},
        title = "{The formation of solar prominences: plasma origin and mechanisms}",
      journal = {Reviews of Modern Plasma Physics},
         year = 2025,
        month = dec,
       volume = {9},
       number = {1},
          eid = {32},
        pages = {32},
          doi = {10.1007/s41614-025-00206-6},
archivePrefix = {arXiv},
       eprint = {2511.14374},
 primaryClass = {astro-ph.SR},
       adsurl = {https://ui.adsabs.harvard.edu/abs/2025RvMPP...9...32Z}
}

@ARTICLE{abramowicz1988,
       author = {{Abramowicz}, M.~A. and {Czerny}, B. and {Lasota}, J.~P. and {Szuszkiewicz}, E.},
        title = "{Slim Accretion Disks}",
      journal = {\apj},
         year = 1988,
        month = sep,
       volume = {332},
        pages = {646},
          doi = {10.1086/166683},
       adsurl = {https://ui.adsabs.harvard.edu/abs/1988ApJ...332..646A}
}

@ARTICLE{chiang2017,
       author = {{Chiang}, Chia-Ying and {Cackett}, E.~M. and {Zoghbi}, A. and {Fabian}, A.~C. and {Kara}, E. and {Parker}, M.~L. and {Reynolds}, C.~S. and {Walton}, D.~J.},
        title = "{X-ray lags in PDS 456 revealed by Suzaku observations}",
      journal = {\mnras},
         year = 2017,
        month = dec,
       volume = {472},
       number = {2},
        pages = {1473-1481},
          doi = {10.1093/mnras/stx2069},
archivePrefix = {arXiv},
       eprint = {1703.08223},
 primaryClass = {astro-ph.HE},
       adsurl = {https://ui.adsabs.harvard.edu/abs/2017MNRAS.472.1473C}
}

@ARTICLE{pounds2025,
       author = {{Pounds}, Ken and {Page}, Kim},
        title = "{Observing the launch of an Eddington wind in the luminous Seyfert galaxy PG1211+143}",
      journal = {\mnras},
         year = 2025,
        month = jul,
       volume = {540},
       number = {3},
        pages = {2530-2534},
          doi = {10.1093/mnras/staf637},
archivePrefix = {arXiv},
       eprint = {2310.18105},
 primaryClass = {astro-ph.HE},
       adsurl = {https://ui.adsabs.harvard.edu/abs/2025MNRAS.540.2530P}
}

@ARTICLE{papaloizou1984,
       author = {{Papaloizou}, J.~C.~B. and {Pringle}, J.~E.},
        title = "{The dynamical stability of differentially rotating discs with constant specific angular momentum}",
      journal = {\mnras},
         year = 1984,
        month = jun,
       volume = {208},
        pages = {721-750},
          doi = {10.1093/mnras/208.4.721},
       adsurl = {https://ui.adsabs.harvard.edu/abs/1984MNRAS.208..721P}
}

@ARTICLE{fabian2012,
       author = {{Fabian}, A.~C.},
        title = "{Observational Evidence of Active Galactic Nuclei Feedback}",
      journal = {\araa},
         year = 2012,
        month = sep,
       volume = {50},
        pages = {455-489},
          doi = {10.1146/annurev-astro-081811-125521},
archivePrefix = {arXiv},
       eprint = {1204.4114},
 primaryClass = {astro-ph.CO},
       adsurl = {https://ui.adsabs.harvard.edu/abs/2012ARA&A..50..455F}
}

@ARTICLE{fabian2020,
       author = {{Fabian}, A.~C. and {Reynolds}, C.~S. and {Jiang}, J. and {Pinto}, C. and {Gallo}, L.~C. and {Parker}, M.~L. and {Lasenby}, A.~N. and {Alston}, W.~N. and {Buisson}, D.~J.~K. and {Cackett}, E.~M. and {De Marco}, B. and {Garcia}, J. and {Kara}, E. and {Kosec}, P. and {Middleton}, M.~J. and {Miller}, J.~M. and {Miniutti}, G. and {Walton}, D.~J. and {Wilkins}, D.~R. and {Young}, A.~J.},
        title = "{Blueshifted absorption lines from X-ray reflection in IRAS 13224-3809}",
      journal = {\mnras},
         year = 2020,
        month = apr,
       volume = {493},
       number = {2},
        pages = {2518-2522},
          doi = {10.1093/mnras/staa482},
archivePrefix = {arXiv},
       eprint = {2002.06388},
 primaryClass = {astro-ph.HE},
       adsurl = {https://ui.adsabs.harvard.edu/abs/2020MNRAS.493.2518F}
}

@ARTICLE{marinucci2020,
       author = {{Marinucci}, A. and {Bianchi}, S. and {Braito}, V. and {De Marco}, B. and {Matt}, G. and {Middei}, R. and {Nardini}, E. and {Reeves}, J.~N.},
        title = "{The lively accretion disc in NGC 2992 - I. Transient iron K emission lines in the high-flux state}",
      journal = {\mnras},
         year = 2020,
        month = aug,
       volume = {496},
       number = {3},
        pages = {3412-3423},
          doi = {10.1093/mnras/staa1683},
archivePrefix = {arXiv},
       eprint = {2006.05280},
 primaryClass = {astro-ph.HE},
       adsurl = {https://ui.adsabs.harvard.edu/abs/2020MNRAS.496.3412M}
}

@ARTICLE{iwasawa2004,
       author = {{Iwasawa}, K. and {Miniutti}, G. and {Fabian}, A.~C.},
        title = "{Flux and energy modulation of redshifted iron emission in NGC 3516: implications for the black hole mass}",
      journal = {\mnras},
         year = 2004,
        month = dec,
       volume = {355},
       number = {4},
        pages = {1073-1079},
          doi = {10.1111/j.1365-2966.2004.08392.x},
archivePrefix = {arXiv},
       eprint = {astro-ph/0409293},
 primaryClass = {astro-ph},
       adsurl = {https://ui.adsabs.harvard.edu/abs/2004MNRAS.355.1073I}
}

@ARTICLE{dauser2010,
       author = {{Dauser}, T. and {Wilms}, J. and {Reynolds}, C.~S. and {Brenneman}, L.~W.},
        title = "{Broad emission lines for a negatively spinning black hole}",
      journal = {\mnras},
         year = 2010,
        month = dec,
       volume = {409},
       number = {4},
        pages = {1534-1540},
          doi = {10.1111/j.1365-2966.2010.17393.x},
archivePrefix = {arXiv},
       eprint = {1007.4937},
 primaryClass = {astro-ph.HE},
       adsurl = {https://ui.adsabs.harvard.edu/abs/2010MNRAS.409.1534D}
}

@ARTICLE{cunningham1975,
       author = {{Cunningham}, C.~T.},
        title = "{The effects of redshifts and focusing on the spectrum of an accretion disk around a Kerr black hole.}",
      journal = {\apj},
         year = 1975,
        month = dec,
       volume = {202},
        pages = {788-802},
          doi = {10.1086/154033},
       adsurl = {https://ui.adsabs.harvard.edu/abs/1975ApJ...202..788C}
}

@ARTICLE{gates2020,
       author = {{Gates}, Delilah E.~A. and {Hadar}, Shahar and {Lupsasca}, Alexandru},
        title = "{Maximum observable blueshift from circular equatorial Kerr orbiters}",
      journal = {\prd},
         year = 2020,
        month = nov,
       volume = {102},
       number = {10},
          eid = {104041},
        pages = {104041},
          doi = {10.1103/PhysRevD.102.104041},
archivePrefix = {arXiv},
       eprint = {2009.03310},
 primaryClass = {gr-qc},
       adsurl = {https://ui.adsabs.harvard.edu/abs/2020PhRvD.102j4041G}
}

@ARTICLE{pds456,
       author = {{XRISM Collaboration} and {Audard}, Marc and {Awaki}, Hisamitsu and {Ballhausen}, Ralf and {Bamba}, Aya and {Behar}, Ehud and {Boissay-Malaquin}, Rozenn and {Brenneman}, Laura and {Brown}, Gregory V. and {Corrales}, Lia and {Costantini}, Elisa and {Cumbee}, Renata and {Trigo}, Mar{\'\i}a D{\'\i}az and {Done}, Chris and {Dotani}, Tadayasu and {Ebisawa}, Ken and {Eckart}, Megan and {Eckert}, Dominique and {Enoto}, Teruaki and {Eguchi}, Satoshi and {Ezoe}, Yuichiro and {Foster}, Adam and {Fujimoto}, Ryuichi and {Fujita}, Yutaka and {Fukazawa}, Yasushi and {Fukushima}, Kotaro and {Furuzawa}, Akihiro and {Gallo}, Luigi and {Garc{\'\i}a}, Javier A. and {Gu}, Liyi and {Guainazzi}, Matteo and {Hagino}, Kouichi and {Hamaguchi}, Kenji and {Hatsukade}, Isamu and {Hayashi}, Katsuhiro and {Hayashi}, Takayuki and {Hell}, Natalie and {Hodges-Kluck}, Edmund and {Hornschemeier}, Ann and {Ichinohe}, Yuto and {Ishida}, Manabu and {Ishikawa}, Kumi and {Ishisaki}, Yoshitaka and {Kaastra}, Jelle and {Kallman}, Timothy and {Kara}, Erin and {Katsuda}, Satoru and {Kanemaru}, Yoshiaki and {Kelley}, Richard and {Kilbourne}, Caroline and {Kitamoto}, Shunji and {Kobayashi}, Shogo and {Kohmura}, Takayoshi and {Kubota}, Aya and {Leutenegger}, Maurice and {Loewenstein}, Michael and {Maeda}, Yoshitomo and {Markevitch}, Maxim and {Matsumoto}, Hironori and {Matsushita}, Kyoko and {McCammon}, Dan and {McNamara}, Brian and {Mernier}, Fran{\c{c}}ois and {Miller}, Eric D. and {Miller}, Jon M. and {Mitsuishi}, Ikuyuki and {Mizumoto}, Misaki and {Mizuno}, Tsunefumi and {Mori}, Koji and {Mukai}, Koji and {Murakami}, Hiroshi and {Mushotzky}, Richard and {Nakajima}, Hiroshi and {Nakazawa}, Kazuhiro and {Ness}, Jan-Uwe and {Nobukawa}, Kumiko and {Nobukawa}, Masayoshi and {Noda}, Hirofumi and {Odaka}, Hirokazu and {Ogawa}, Shoji and {Ogorzalek}, Anna and {Okajima}, Takashi and {Ota}, Naomi and {Paltani}, Stephane and {Petre}, Robert and {Plucinsky}, Paul and {Porter}, Frederick Scott and {Pottschmidt}, Katja and {Sato}, Kosuke and {Sato}, Toshiki and {Sawada}, Makoto and {Seta}, Hiromi and {Shidatsu}, Megumi and {Simionescu}, Aurora and {Smith}, Randall and {Suzuki}, Hiromasa and {Szymkowiak}, Andrew and {Takahashi}, Hiromitsu and {Takeo}, Mai and {Tamagawa}, Toru and {Tamura}, Keisuke and {Tanaka}, Takaaki and {Tanimoto}, Atsushi and {Tashiro}, Makoto and {Terada}, Yukikatsu and {Terashima}, Yuichi and {Tsuboi}, Yohko and {Tsujimoto}, Masahiro and {Tsunemi}, Hiroshi and {Tsuru}, Takeshi G. and {Uchida}, Hiroyuki and {Uchida}, Nagomi and {Uchida}, Yuusuke and {Uchiyama}, Hideki and {Ueda}, Yoshihiro and {Uno}, Shinichiro and {Vink}, Jacco and {Watanabe}, Shin and {Williams}, Brian J. and {Yamada}, Satoshi and {Yamada}, Shinya and {Yamaguchi}, Hiroya and {Yamaoka}, Kazutaka and {Yamasaki}, Noriko and {Yamauchi}, Makoto and {Yamauchi}, Shigeo and {Yaqoob}, Tahir and {Yoneyama}, Tomokage and {Yoshida}, Tessei and {Yukita}, Mihoko and {Zhuravleva}, Irina and {Braito}, Valentina and {Cond{\`o}}, Pierpaolo and {Fukumura}, Keigo and {Gonzalez}, Adam and {Luminari}, Alfredo and {Miyamoto}, Aiko and {Mizukawa}, Ryuki and {Reeves}, James and {Sato}, Riki and {Tombesi}, Francesco and {Xu}, Yerong},
        title = "{Structured ionized winds shooting out from a quasar at relativistic speeds}",
      journal = {\nat},
         year = 2025,
        month = may,
       volume = {641},
       number = {8065},
        pages = {1132-1136},
          doi = {10.1038/s41586-025-08968-2},
archivePrefix = {arXiv},
       eprint = {2505.09171},
 primaryClass = {astro-ph.HE},
       adsurl = {https://ui.adsabs.harvard.edu/abs/2025Natur.641.1132X}
}

@ARTICLE{dadina2026,
       author = {{Dadina}, M. and {Missaglia}, V. and {Braito}, V. and {Cappi}, M. and {Luminari}, A. and {Barret}, D. and {Bertola}, E. and {Bianchi}, S. and {Comastri}, A. and {Chartas}, G. and {Kaastra}, J. and {Kammoun}, E. and {Lanzuisi}, G. and {Matzeu}, G. and {Middei}, R. and {Nardini}, E. and {Nicastro}, F. and {Petrucci}, Pierre-Olivier and {Pinto}, C. and {Serafinelli}, R. and {Tortosa}, A. and {Vignali}, C.},
        title = "{Is XRISM/Resolve probing a ``raining'' absorber in Mrk 509?}",
      journal = {arXiv e-prints},
         year = 2026,
        month = may,
          eid = {arXiv:2605.02662},
        pages = {arXiv:2605.02662},
          doi = {10.48550/arXiv.2605.02662},
archivePrefix = {arXiv},
       eprint = {2605.02662},
 primaryClass = {astro-ph.HE},
       adsurl = {https://ui.adsabs.harvard.edu/abs/2026arXiv260502662D}
}

@ARTICLE{reeves2000,
       author = {{Reeves}, J.~N. and {O'Brien}, P.~T. and {Vaughan}, S. and {Law-Green}, D. and {Ward}, M. and {Simpson}, C. and {Pounds}, K.~A. and {Edelson}, R.},
        title = "{PDS 456: an extreme accretion rate quasar?}",
      journal = {\mnras},
         year = 2000,
        month = feb,
       volume = {312},
       number = {2},
        pages = {L17-L21},
          doi = {10.1046/j.1365-8711.2000.03282.x},
archivePrefix = {arXiv},
       eprint = {astro-ph/9911475},
 primaryClass = {astro-ph},
       adsurl = {https://ui.adsabs.harvard.edu/abs/2000MNRAS.312L..17R}
}

@ARTICLE{torres1997,
       author = {{Torres}, Carlos A.~O. and {Quast}, Germano R. and {Coziol}, Roger and {Jablonski}, Francisco and {de la Reza}, Ramiro and {L{\'e}pine}, J.~R.~D. and {Greg{\'o}rio-Hetem}, J.},
        title = "{Discovery of a Luminous Quasar in the Nearby Universe}",
      journal = {\apjl},
         year = 1997,
        month = oct,
       volume = {488},
       number = {1},
        pages = {L19-L22},
          doi = {10.1086/310913},
archivePrefix = {arXiv},
       eprint = {astro-ph/9707258},
 primaryClass = {astro-ph},
       adsurl = {https://ui.adsabs.harvard.edu/abs/1997ApJ...488L..19T}
}

@ARTICLE{parker1966,
       author = {{Parker}, E.~N.},
        title = "{The Dynamical State of the Interstellar Gas and Field}",
      journal = {\apj},
         year = 1966,
        month = sep,
       volume = {145},
        pages = {811},
          doi = {10.1086/148828},
       adsurl = {https://ui.adsabs.harvard.edu/abs/1966ApJ...145..811P}
}

@ARTICLE{reynolds2021,
       author = {{Reynolds}, Christopher S.},
        title = "{Observational Constraints on Black Hole Spin}",
      journal = {\araa},
         year = 2021,
        month = sep,
       volume = {59},
        pages = {117-154},
          doi = {10.1146/annurev-astro-112420-035022},
archivePrefix = {arXiv},
       eprint = {2011.08948},
 primaryClass = {astro-ph.HE},
       adsurl = {https://ui.adsabs.harvard.edu/abs/2021ARA&A..59..117R}
}

@ARTICLE{li2024,
       author = {{Li}, Yan-Rong and {Hu}, Chen and {Yao}, Zhu-Heng and {Chen}, Yong-Jie and {Bai}, Hua-Rui and {Yang}, Sen and {Du}, Pu and {Fang}, Feng-Na and {Fu}, Yi-Xin and {Liu}, Jun-Rong and {Peng}, Yue-Chang and {Songsheng}, Yu-Yang and {Wang}, Yi-Lin and {Xiao}, Ming and {Zhai}, Shuo and {Winkler}, Hartmut and {Bai}, Jin-Ming and {Ho}, Luis C. and {Petrov}, Romain G. and {Aceituno}, Jes{\'u}s and {Wang}, Jian-Min},
        title = "{Spectroastrometry and Reverberation Mapping of Active Galactic Nuclei. I. The H{\ensuremath{\beta}} Broad-line Region Structure and Black Hole Masses of Five Quasars}",
      journal = {\apj},
         year = 2024,
        month = oct,
       volume = {974},
       number = {1},
          eid = {86},
        pages = {86},
          doi = {10.3847/1538-4357/ad6906},
archivePrefix = {arXiv},
       eprint = {2407.08120},
 primaryClass = {astro-ph.GA},
       adsurl = {https://ui.adsabs.harvard.edu/abs/2024ApJ...974...86L}
}

@ARTICLE{xu2025,
       author = {{Xu}, Yerong and {Gallo}, Luigi C. and {Hagino}, Kouichi and {Reeves}, James N. and {Tombesi}, Francesco and {Mizumoto}, Misaki and {Luminari}, Alfredo and {Gonzalez}, Adam G. and {Behar}, Ehud and {Boissay-Malaquin}, Rozenn and {Braito}, Valentina and {Cond{\'o}}, Pierpaolo and {Done}, Chris and {Miyamoto}, Aiko and {Mizukawa}, Ryuki and {Odaka}, Hirokazu and {Sato}, Riki and {Tanimoto}, Atsushi and {Tashiro}, Makoto and {Yaqoob}, Tahir and {Yamada}, Satoshi},
        title = "{Unraveling the structure of the stratified ultra-fast outflows in PDS 456 with XRISM}",
      journal = {\pasj},
         year = 2025,
        month = sep,
       volume = {77},
        pages = {S223-S241},
          doi = {10.1093/pasj/psaf070},
archivePrefix = {arXiv},
       eprint = {2506.05273},
 primaryClass = {astro-ph.HE},
       adsurl = {https://ui.adsabs.harvard.edu/abs/2025PASJ...77S.223X}
}




\appendix
\section{Modeling the broad emission feature in the XRISM Resolve spectrum}
\label{app:modeling}
In order to determine the inclination of the accretion disc from the broad emission feature in the iron K band, we apply the model of relativistically blurred reflection from the inner accretion disc to the \textit{XRISM} observation obtained on 2024 March 10 (Observation ID 300072010), with an expsoure time of 264\,ks. The high-resolution spectrum obtained by the \textit{Resolve} microcalorimeter spectrometer was obtained following the standard data reduction steps outlined in the \textit{XRISM ABC Guide}.

The count rate seen from PDS\,456 by \textit{Resolve} is sufficiently low that the vast majority of events were classified as high spectral resolution (`Hp') events. We therefore restrict our analysis to just these events. The spectrum was extracted using \textsc{xselect} from the entire $6\times 6$ pixel array, excluding pixel 27 which shows anomalous gain variations, and pixel 12 (the calibration pixel). The response matrix and ancillary response for this specific observation and event list were generated using the tools \textsc{rslmkrmf} and \textsc{xaarfgen}, respectively.

The model was fit to the spectrum across the 2-12\,keV energy band using \textsc{xspec}, minimizing the C-statistic to find the model parameters that best describe the data. Both the continuum and relativistically blurred reflection from the accretion disc were provided by the \textsc{relxilllpCp} model, which, in calculating the radial emissivity profile of the reflection observed from the disc, assumes illumination by a compact point source located on the spin axis above the black hole. In addition to the inclination fo the accretion disc to the line of sight, free model parameters include the photon index of the continuum, the spin of the black hole, the height of the point source, the iron abundance in the disc and the ionization parameter of the plasma, and the reflection fraction (the ratio of the reflection to directly observed continuum emission). Both the primary continuum and reflection pass through the absorbers. We employ a phenomenological description of the absorption line forest, consisting of five photoionized absorbers, modeled using \textsc{xstar}. The velocity and column density of each is fit to the observed spectrum. The ionization parameter and turbulent velocity broadening are assumed to be the same for all five of the absorption zones. These values are fit to the data and tied between the components.

In addition to the source model, it is necessary to include a model of the instrumental non-X-ray background. This model component includes the background continuum, described by a power law, in addition to instrumental emission lines, and was constructed from the \textit{XRISM Resolve} \textsc{nxb database} for the observation in question, and was added to the model through a diagonal rersponse matrix, following the standard procedure in \textit{xspec}.

The \textit{XRISM} observation was obtained alongisde observations with \textit{NuSTAR}, \textit{XMM-Newton}, \textit{NICER} and \textit{Swift}. Significant variability was detected over the course of the observations, with a flare observed at the beginning. While overlapping, the observations were not strictly simultaneous, meaning that joint analysis of the spectra must account for any variation in the spectrum over the course of the observations. Since we use this model only to place constraints on the inclination of the accretion disc, we therefore fit only the \textit{XRISM Resolve} spectrum in this work for simplicity, marginalizing over any uncertainty in the continuum that arises from fitting the model only to a limited bandpass.

The best-fitting model parameters are shown in Table~\ref{tab:par}. Uncertainties and confidence intervals for each parameter were estimated from Markov chain Monte Carlo calculations, run using the Goodman-Weare algorithm, implemented in \textsc{xspec}, with 100 walkers for $10^7$ steps. To erase any memory of the starting point of the chains, the first 150,000 steps were discarded or `burned'.

\begin{table*}
\caption{The best-fitting parameters of the inner disc reflection model to the \textit{XRISM Resolve} spectrum of PDS\,456. Both the continuum and reflection from the disc are described by the \textsc{relxilllpcp} model, which assumes a simplified point source geometry for the X-ray emitting corona that illuminates the disc. A phenomenological description of the blueshifted absorbers is included, consisting of five photoionized absorption zones. The column density and line-of-sight velocity of each is fit to the data, while the ionization parameter and turbulent broadening are tied between the zones. The redshift of PDS\,456 is taken to be $z = 0.184$. The quoted uncertainties correspond to the $1\sigma$ confidence limit. Upper and lower limits correspond to the 95 per cent interval.}
\label{tab:par}
    \begin{tabular}{llc}
    \hline
    Component & Parameter & Value \\
    \hline
    Absorbers & Ionization parameter, $\log(\xi~/~\mathrm{erg\,cm\,s}^{-1})$ & $4.10_{-0.07}^{+0.17}$ \\
    & Velocity broadening, $\sigma_v~/~\mathrm{km\,s}^{-1}$ & $3380_{-400}^{+70}$ \\
    Zone 1 & Column density, $N_\mathrm{H}~/~10^{22}\,\mathrm{cm}^{-2}$ & $2.3_{-0.9}^{+0.8}$ \\
    & LoS velocity, $v/c$ & $-0.241_{-0.02}^{+0.001}$ \\
    Zone 2 & Column density, $N_\mathrm{H}~/~10^{22}\,\mathrm{cm}^{-2}$ & $1.3_{-0.2}^{+0.6}$ \\
    & LoS velocity, $v/c$ & $-0.253_{-0.009}^{+0.012}$ \\
    Zone 3 & Column density, $N_\mathrm{H}~/~10^{22}\,\mathrm{cm}^{-2}$ & $3.0_{-1.4}^{+0.5}$ \\
    & LoS velocity, $v/c$ & $-0.290_{-0.04}^{+0.001}$ \\
    Zone 4 & Column density, $N_\mathrm{H}~/~10^{22}\,\mathrm{cm}^{-2}$ & $2.2_{-0.5}^{+1.1}$ \\
    & LoS velocity, $v/c$ & $-0.314_{-0.001}^{+0.03}$ \\
    Zone 5 & Column density, $N_\mathrm{H}~/~10^{22}\,\mathrm{cm}^{-2}$ & $2.9_{-1.1}^{+0.4}$ \\
    & LoS velocity, $v/c$ & $-0.344_{-0.016}^{+0.013}$ \\
    \hline
    Continuum and disc reflection & Photon index, $\Gamma$ & $2.17_{-0.13}^{+0.05}$ \\
    & Reflection fraction, $R_F$ & $2.4_{-0.1}^{+1.0}$ \\
    & Corona height, $h~/~r_\mathrm{g}$ & $7_{-1}^{+1}$ \\
    & Inclination, $i~/~\mathrm{deg}$ & $70.1_{-0.4}^{+2.6}$ \\
    & Iron abundance, $A_\mathrm{Fe}~/~\mathrm{Solar}$ & $>4$ \\
    & Ionization parameter, $\log(\xi~/~\mathrm{erg\,cm\,s}^{-1})$ & $<0.1$ \\
    \hline
    \end{tabular}
\end{table*}


\bsp	
\label{lastpage}
\end{document}